\documentclass[smallextended]{svjour3} 
\pdfoutput=1
\usepackage{amsmath}
\usepackage{amssymb}
\usepackage{graphicx}
\usepackage{float}
\usepackage{url}
\usepackage{hyperref}

\def\b{\beta}

\newcommand{\E}{\ensuremath{\mathbb{E}}}

\def\be{\begin{equation}}
\def\ee{\end{equation}}
\def\beq{\begin{array}}
\def\eeq{\end{array}}

\begin{document}
\title{Inferring the COVID-19 infection curve in Italy}
\author{Andrea Pugliese \and Sara Sottile}
\institute{A. Pugliese \at Dipartimento di Matematica, Universit\`a di Trento\\ \email{andrea.pugliese@unitn.it} 
\and S. Sottile \at Dipartimento di Matematica, Universit\`a di Trento\\ \email{sara.sottile@unitn.it} }
\date{}
\maketitle
\begin{abstract}
Aim of this manuscript is to show a simple method to infer the time-course of new COVID-19 infections (the most important information in order to establish the effect of containment strategies) from available aggregated data, such as number of deaths and hospitalizations. The method, that was used for HIV-AIDS and was named `back-calculation', relies on good estimates of the distribution of the delays between infection and the observed events; assuming that the epidemic follows a simple SIR model with a known generation interval, we can then estimate the parameters that define the time-varying contact rate through maximum likelihood. We show the application of the method to data from Italy and several of its region; it is found that $R_0$ had decreased consistently below 1 around March 20, and in the beginning of April it was between 0.5 and 0.8 in the whole Italy and in most regions.
\end{abstract}

\section{Introduction}
The emergence of infections from SARS-CoV-2  first in China and then in many other countries of the world has, at the moment, caused more than 100,000 deaths all over the world, and very severe disease  in a large number of people. Moreover, while the response to the epidemic with `lock-down' strategies has probably prevented a much larger number of deaths and saved from dire consequences, it is also causing huge economic losses that can affect the well-being of millions of people.

Thus most countries experiencing the lock-down have already started or are planning re-opening strategies that could allow for a socio-economic recovery while maintaining the infections under control. 
Such strategies will have rely on many factors: adequate planning of activities, in order to maintain physical distances; large availability of testing facilities; technological tools that help in tracking contacts of infected individuals; collaboration of citizens in adhering to the recommendations of health authorities.  
In order to be able to handle appropriately the phase of partial re-opening, it will also be crucial to have reliable  and timely estimates of the trend in new infections and in the reproduction number $R_0$.

Here we show how it is possible to obtain such estimates using publicly available data on deaths and hospitalizations due to COVID-19 infections, and estimates of the delays between infections and such events. Clearly, if one  obtains reliable data from the health system or digital tools on many other events, such as symptom onset or infection diagnosis, the method would yield more precise and timely estimates.
 The method shown here could be adapted to many different information, as long as estimates exist of  the distribution of the time between infection  and the event recorded.
 
 The idea is very simple and was introduced in epidemiology at the time of the emergence of HIV/AIDS: knowing the number of patients that during a certain period seeked care for symptoms related to AIDS syndrome, one would attempt, moving backwards in time for the length of the incubation period,  to `back-calculate' the curve of HIV infections \cite{Brookmeyer1988}.\\
  An important difference between HIV/AIDS and the COVID-19 infection, beyond the big difference in the time-scales involved, is that 
 it was generally agreed that (almost) all patients infected with HIV would develop AIDS; hence, in principle, from the data on AIDS cases one could estimate the actual curve of HIV infections of several years earlier. On the other hand, it is known that a large proportion of infections with SARS-CoV2 are asymptomatic or mild and thus not diagnosed and not reported. \\
 Hence from the reported cases, one cannot recover the number of all infections, without other information. The output of the method will be an estimated curve of the infection times of the reported cases. Assuming that the proportion of infections that are reported is constant in time, the curve of actual infections will be a multiple of the estimated curve.
 
Italy was the first European country with a significant spread of COVID-19 infection, and has been struck very severely with more than 20,000 deaths. We apply this method to the Italian data published every day from the Protezione Civile.  The method bears strong similarities with what has been performed by Flaxman \textit{et al.} \cite{Flaxman2020} for several European countries, including Italy. In the Discussion we will better explain the differences and similarities between our approach and \cite{Flaxman2020}.

Italian data on infections from SARS-CoV-2 have been recently analysed by \cite{Sebastiani2020}. Their method of analysis was instead very different from what used here.
 
 \section{Data and information used}
 In this manuscript we use the data concerning COVID-19 infections in  Italy daily published by the Protezione Civile \cite{covidItalia}. The tables report, for each region (NUTS-2 subdivision level) of Italy, the number of daily new reported cases; the cumulative numbers of dead, of discharged patients, and of swabs performed; the current numbers of hospitalized patients, of patients in ICUs, or in home isolation. We decided that the number of reported cases is not indicative of the actual number of infected people, as the number depends strongly on the testing policy, which has widely varied during the epidemic. We chose to use instead the number of daily deaths and hospitalizations as useful information to infer the curve of new infections. 
 
 As mentioned, the reported data include  $TD_t$ (total deaths up to day $t$) and $CH_t$ (currently hospitalized patients), while our analysis is based on $D_t$ (deaths in day $t$) and $H_t$ (new hospitalizations in day $t$). We used the following estimates
\begin{equation}
\label{data}
 D_t = TD_t - TD_{t-1} \qquad \mbox{and}\quad H_t = CH_t - CH_{t-1} + p D_t + (TR_t -TR_{t-1}). 
\end{equation}
The first formula is obvious. The second one starts by taking new hospitalizations as the difference between hospitalized of today and of yesterday; we have however to add all patients that were hospitalized yesterday and are no longer in the hospital: those are the discharged because recovered ($TR_t -TR_{t-1}$, where $TR_t$ is the reported cumulative number of discharged) and the dead ones: it is assumed that a proportion $p$ of today's dead were in hospital the previous day. In most analysis, we will use $p=0.8$, but the results are almost identical setting $p=1$.

Finally, since data sometimes vary widely from one day to the next (possibly because of reporting delays), we regularize them through a 3-day mobile media in the past, Namely, we use $h_t = \frac{1}{ 3}\left(H_t + H_{t-1}+H_{t-2}\right)$  and similarly $d_t =  \frac{1}{ 3}\left(D_t + D_{t-1}+D_{t-2}\right)$.

Having decided to use as informative events hospitalizations and deaths, the method is based on knowing the distribution of the time between infection and hospitalization; and of the time between infection and death.
We could not find in the literature direct estimates of these delays.

On the other hand, there are several estimates of the incubation period (time between infection and symptom onset), and some of the intervals between symptom onset and hospitalization or death. 
Assuming that the length of the incubation period is independent of the time spent between symptom onset and hospitalization or death, we can then obtain the required distribution by a simple convolution.

\begin{figure}[H]
\begin{center}
\includegraphics[width=6.5cm]{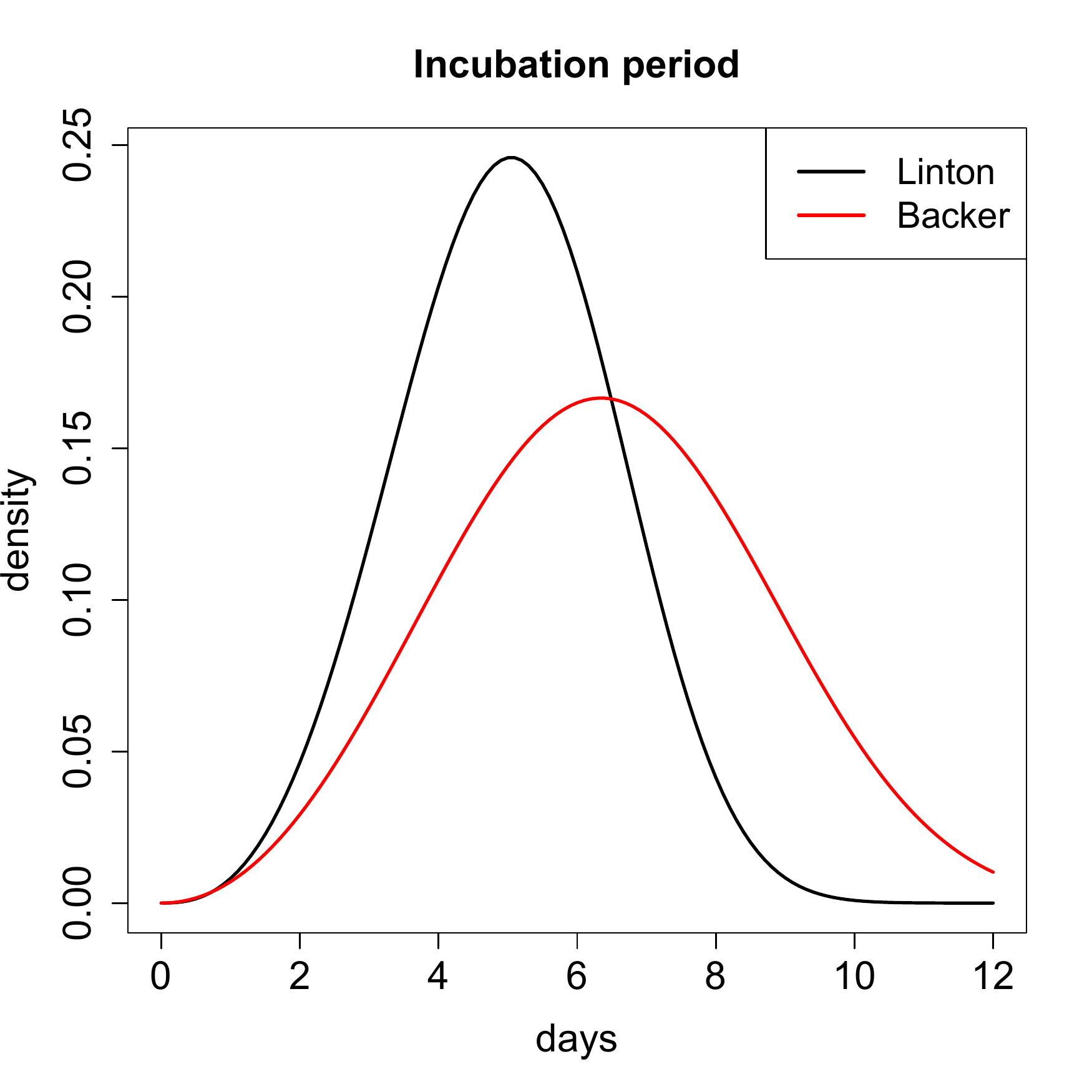}\ \includegraphics[width=6.5cm]{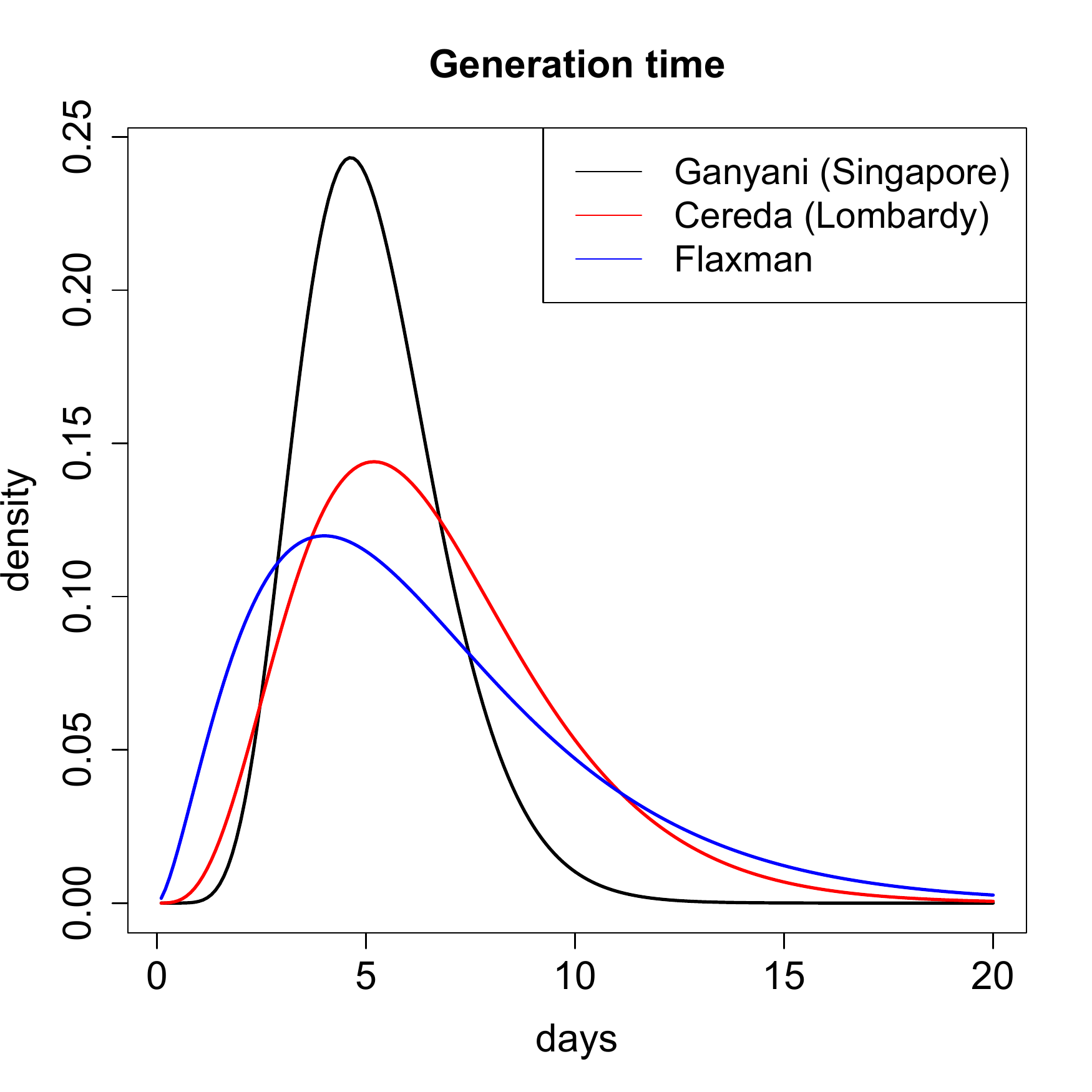}
\caption{Left: different estimates of the incubation period. Right: different estimates of the generation time,}
\label{fig:inc}
\end{center}
\end{figure}

Of the several existing estimates of the incubation period, we singled out those obtained by Linton \textit{et al.} \cite{Linton2020} 
 and by Backer \textit{et al.} \cite{Backer2020}
shown in Fig.~\ref{fig:inc}a.

Another important component of the method shown in the next Section is the distribution of the generation time, i.e. the interval between the infection time of an individual and the infection time of an individual infected by the former.

We found several estimates also concerning the generation time,.
The ones that have been used in our analysis are the one estimated by Ganyani \textit{et al.} \cite{Ganyani2020}
using data from Singapore; the one estimated by Cereda \textit{et al.} analysing early cases in Lombardy \cite{Cereda2020},
 and that proposed by Flaxman \textit{et al.} \cite{Flaxman2020}
 The three distributions are shown in Fig.~\ref{fig:inc}b. Clearly, those by Cereda \textit{et al.} \cite{Cereda2020} and by Flaxman \textit{et al.} \cite{Flaxman2020} have a much heavier right tail than the one estimated by Ganyani \textit{et al.} \cite{Ganyani2020}. As the length of generation time is influenced by the social context, it is possible that in Singapore the authoririties were very efficient in identifying and isolating infected individuals, so as to avoid long periods of infectivity. We thus decided to use only the other ones.

\begin{figure}[htb]
\begin{center}
\includegraphics[width=6.5cm]{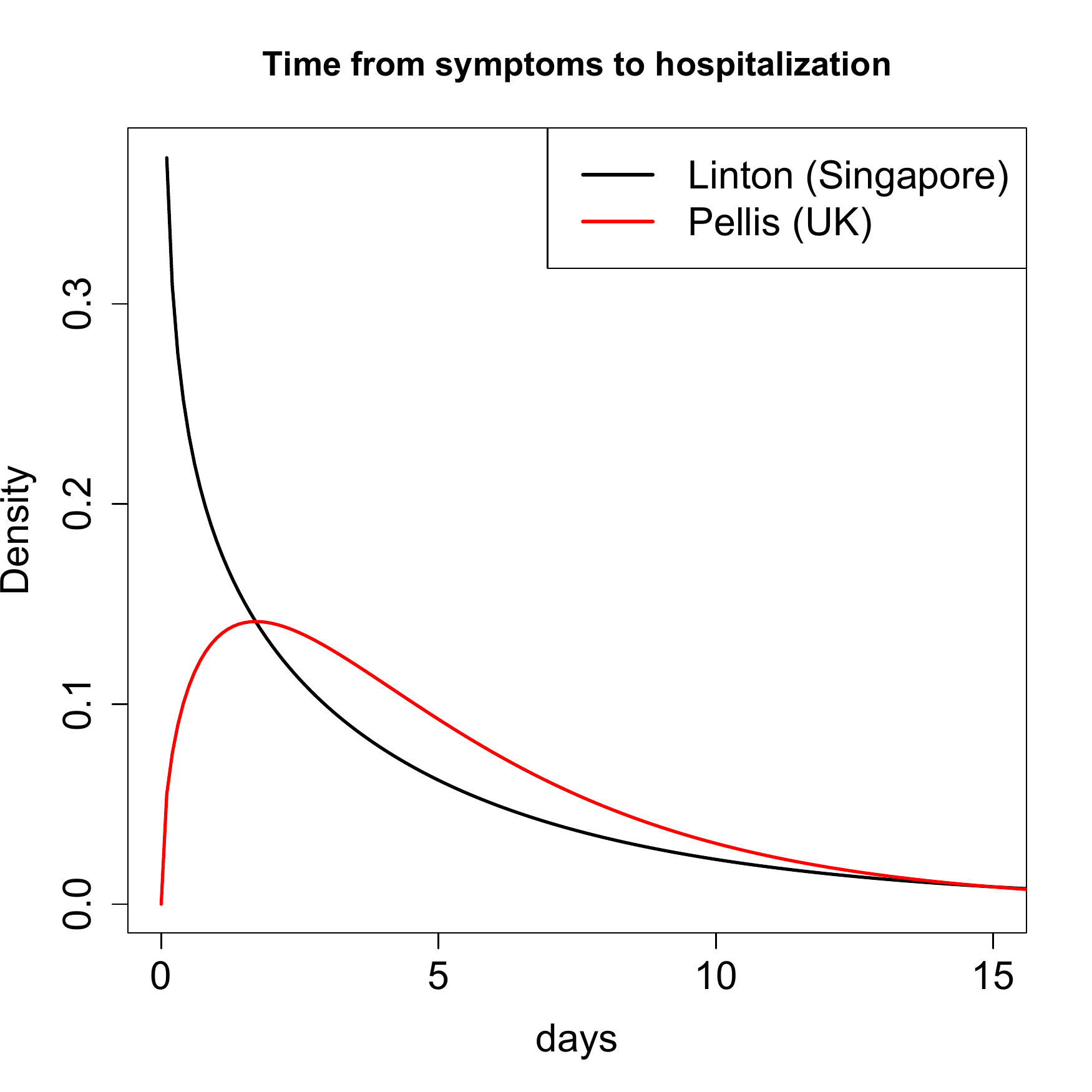}\ \includegraphics[width=6.5cm]{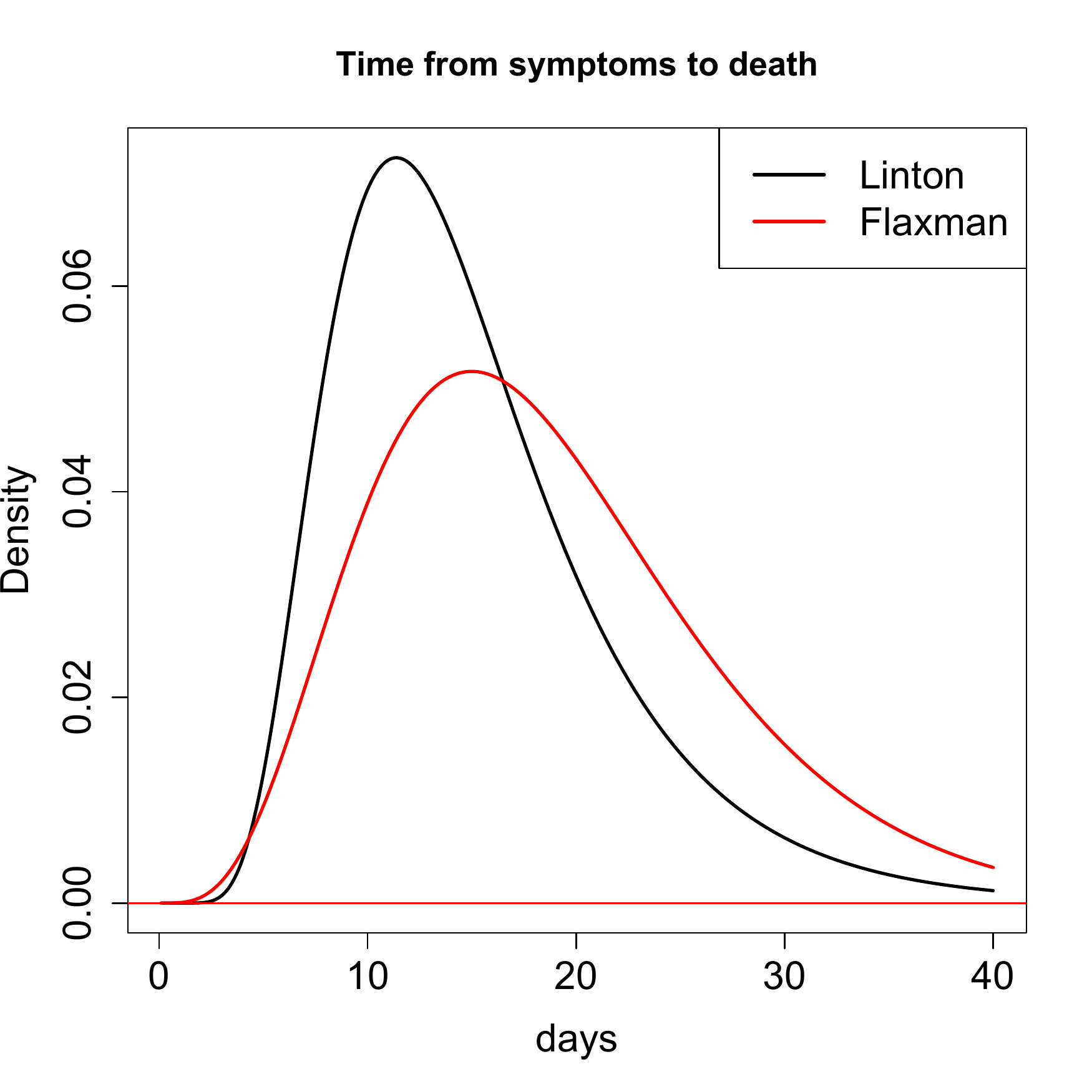}
\caption{Left: estimates of the time from symptom onset to hospitalization. Right: estimates of the time from symptom onset to death.}
\label{fig:hosp_death}
\end{center}
\end{figure}

We found fewer estimates for the times from symptom onset to hospitalization or to death. Concerning the former, we found an estimate by Linton \textit{et al.} \cite{Linton2020} based on data from Singapore, and one by Pellis \textit{et al.} \cite{Pellis2020} based on data from UK, which are shown in Fig.~\ref{fig:hosp_death}a. Linton \textit{et al.}  \cite{Linton2020} provide also an estimate of the distribution of the time from symptom onset to death; another estimate is proposed by Flaxman \textit{et al.} \cite{Flaxman2020} elaborating on estimates of severity \cite{Verity2020}; the two distributions are shown in Fig.~\ref{fig:hosp_death}b.

\begin{figure}[htb]
\begin{center}
\includegraphics[width=6.5cm]{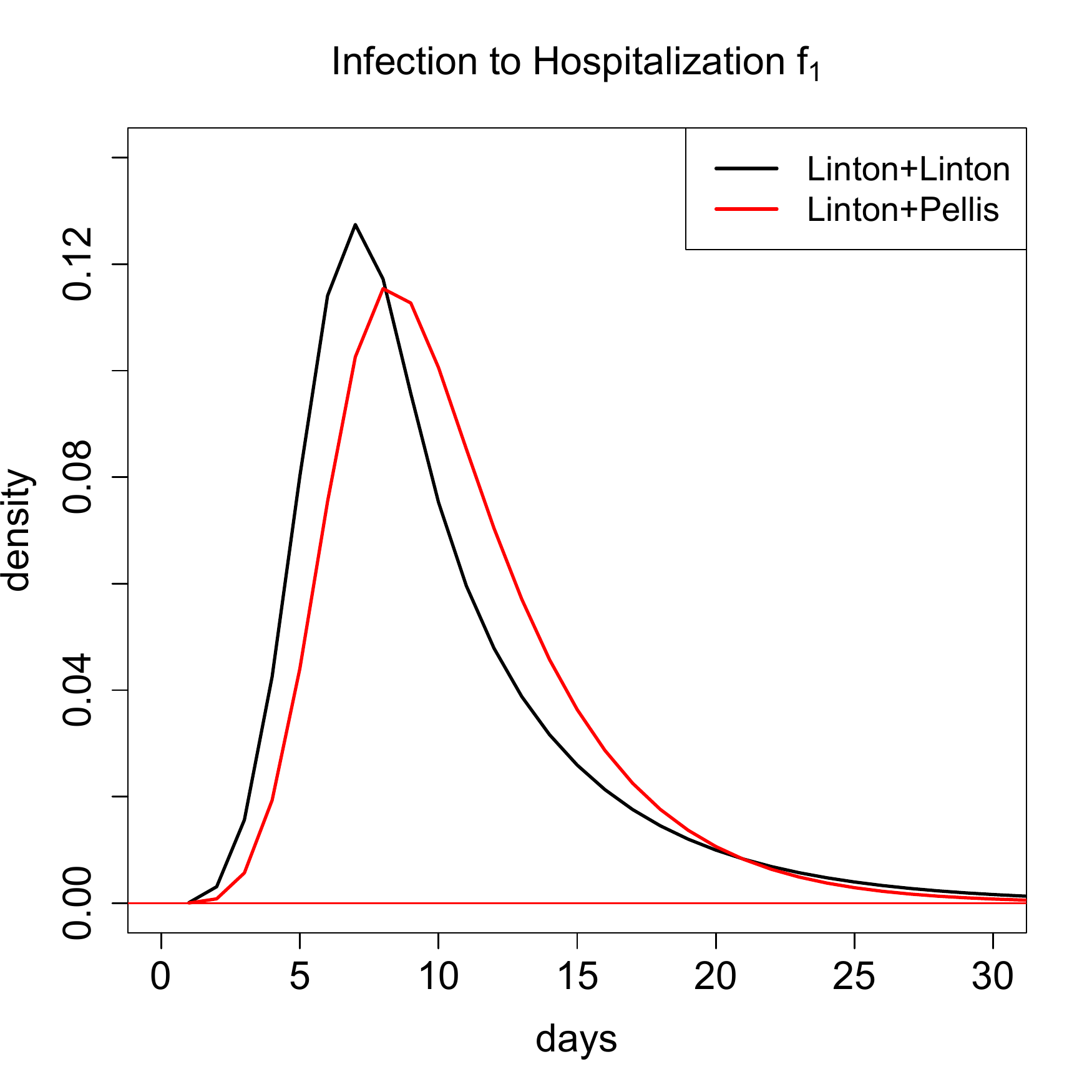}\ \includegraphics[width=6.5cm]{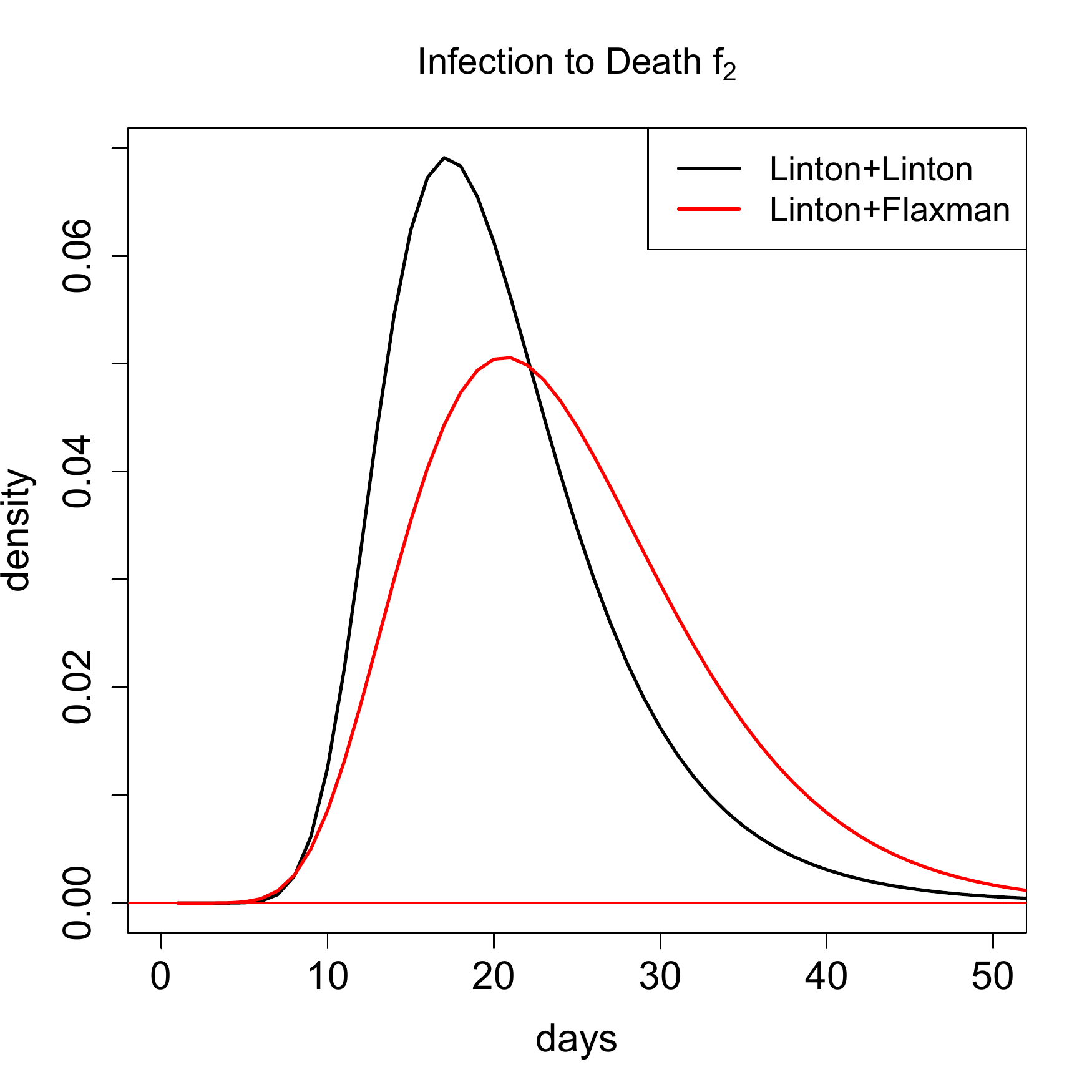}
\caption{Left: estimates of the time from infection to hospitalization. Right: estimates of the time from infection to death.}
\label{fig:f1}
\end{center}
\end{figure}

From the convolution of the density of the incubation period with the densities of the time from symptom onset to hospitalization or death, we obtain $f_1(t)$ (Fig.~\ref{fig:f1}a), density of the time from infection to hospitalization, and  $f_2(t)$ (Fig.~\ref{fig:f1}b), density of the time from infection to death. Fig.~\ref{fig:f1} only shows the values obtained, using the incubation period estimated by Linton \textit{et al.} \cite{Linton2020}, as that has been mainly used in the analysis.
  \section{Methods}
 Suppose that we know the density $f$ of the time between infection and an event that is recorded. Letting $E(t)$ rate of event recording and $j(t)$ infection incidence, and assuming that infections have started at time $t=0$, we have
 \begin{equation}
\label{eq_base}
E(t) =  p \int_0^t f(t-s) j(s) \, ds
\end{equation}
where $p$ is the probability that an infection gives rise (sooner or later) to a recorded event.\\
If $E(t)$ is a known function, \eqref{eq_base} is a Volterra integral equation of the 1st kind  in the unknown $p\,j(t)$; under minimal assumptions, such equation has  a unique solution, so that in principle $p\,j(t)$ could be recovered exactly.\\
However, first of all,  $E(t)$ is reported at discrete intervals, daily in the case of COVID-19 infections. Hence, letting $E_t$ the number of events recorded in day $t$, \eqref{eq_base} becomes
\begin{multline} 
\label{eq1}
E_t = p \int_{t-1}^t  \int_0^s f(s-u) j(u) \, du \, ds \\=p \int_0^{t-1} j(u) \int_{t-1}^t f(s-u) \, ds\, du +p \int_{t-1}^{t} j(u) \int_{s}^t f(s-u) \, ds\, du \\
\approx p\int_0^{t-1} j(u)  f(t-\frac{1}{2 }-u) du .
\end{multline}
The second integral in the second line of \eqref{eq1} has been neglected, since typically $f(u) \approx 0$ for $u < 1$, while the inner integral in the first term is approximated with the midpoint rule. If needed, more accurate approximations could be used, but the approximation in \eqref{eq1} seems adequate for our aims, since many uncertainties exist on other aspects of the problem.

While, as stated above, there exists a unique solution $p\, j(\cdot)$ of \eqref{eq_base}, such a problem is notoriouly ill-posed; thus, even small errors (or discretizations) in the functions $E(t)$ or $f(u)$ could give rise to  computed solutions completely different from the true solution. Regularization methods have been proposed to obtain approximate solutions from a purely analytical approach \cite{Lamm1996}. 

Following the approach generally used in epidemiology \cite{Wild1993}, we interpret instead \eqref{eq_base} from a probabilistic point of view, and we use parametric inference assuming that the incidence function $j(t)$ results from a standard epidemiological model.

Precisely, we assume that $j(t)$ is the solution of 
\begin{equation}
\label{Volterra}
j(t) = \beta(t) \int_0^t \phi(s) j(t-s)\, ds + \beta(t) j_0(t).
\end{equation}
Equation \eqref{Volterra} is a simple modification of the model presented by Kermack and McKendrick \cite{kermack_mckendrick} (see, e.g., Diekmann \textit{et al.} \cite{Diekmann_bk2} for a more modern presentation and Iannelli-Milner \cite{IannelliMilner} for a detailed analysis) 
where $j_0(t)$ represents the infectivity at time $t$ of the infectives already present at $t=0$ or introduced from outside during the epidemic, $\phi(\cdot)$ is the infectivity kernel (density of generation time in usual epidemiological terminology) and $\beta(t)$ the time-varying contact rate.
The susceptible fraction does not appear in the equation, because we expect that, despite the high burden caused by the COVID-19 epidemic, the reduction of the susceptible fraction has been very modest in these months, except perhaps in a few small hot spots. However, a possible reduction in susceptibles would be interpreted in \eqref{Volterra} as a reduction of contact rate.

The two events that will be considered in this manuscript are deaths and hospitalizations due to COVID-19. Letting $H_t$ the new hospitalizations, and $D_t$ the deaths reported in day $t$, we assume that these are independent random variable, whose mean is given by \eqref{eq1} for an appropriate delay density $f$. \\
Precisely we assume that either
\begin{equation}
\label{Pois}
\begin{split}
H_t & \sim \mbox{\it Poisson}\left(p_1\int_0^{t-1} j(u)  f_1(t-\frac{1}{2 }-u) du\right) \\ D_t &\sim \mbox{\it Poisson}\left(p_2\int_0^{t-1} j(u)  f_2(t-\frac{1}{2 }-u) du\right)
\end{split}
\end{equation}
or
\begin{equation}
\label{negbin}
\begin{split}
H_t &\sim \mbox{\it NegBin}\left( \mu = p_1\int_0^{t-1} j(u)  f_1(t-\frac{1}{2 }-u) du,\ k\right) \\ D_t &\sim \mbox{\it NegBin}\left(\mu = p_2\int_0^{t-1} j(u)  f_2(t-\frac{1}{2 }-u) du,\ k\right).
\end{split}
\end{equation}
In \eqref{Pois} or \eqref{negbin}, $j(\cdot)$ is the unique solution of \eqref{Volterra} where $\phi(\cdot)$ is taken as a known function, $j_0(\cdot)$ known up to a multiplicative function $C$ and $\beta(\cdot)$ will be a function parameterised by some parameters $(\b_1,\ldots, \b_m)$. Thus 
$ j(u) = C j(u; \b_1,\ldots, \b_m). $ Furthermore, $f_1(\cdot)$ and $f_2(\cdot)$ are taken as known functions, while $p_1$ and $p_2$ are constants to be estimated. Finally, when the negative binomial \eqref{negbin} is used,  we take, for the sake of simplicity, $k$, the dispersion parameter, as known (see below).

Then, with either \eqref{Pois} or \eqref{negbin}, we have that $\E(H_t) = C_1 G(\b_1,\ldots, \b_m)$ and $\E(D_t) = C_2 G(\b_1,\ldots, \b_m)$, where $C_1 = C p_1$ and $C_2 = C p_2$. From the  observed data $\{h_1,\ldots, h_n\}$ and $\{d_1,\ldots,d_n\}$ at the times $t_i = t_0 + i$, $i=1\ldots n$, we obtain a pseudo-likelihood function
\begin{equation}
\label{likelihood}
L(C_1,C_2,\b_1,\ldots, \b_m) = \left(\prod_{i=1}^n P(H_{t_i} = h_i) \right)^{q} \left(\prod_{i=1}^n P(H_{t_i} = h_i) \right)^{1- q}
\end{equation}
where $q$ is a parameter between $0$ and $1$. When $q=0$, it is the actual likelihood based on the data on dead patients, and $L$ does not depend on $C_1$; symmetrically, if $q=1$, it is the likelihood based on the hospitalization data, and $L$ does not depend on $C_2$; finally when $q=1/2$, maximizing $L$ is equivalent to maximing the true likelihood under the assumption of independence of $H_t$ and $D_t$.

The functions $\beta(\cdot)$ used in \eqref{Volterra} have been chosen as cubic spline functions with knots at times $T_1 < T_2 < \cdots < T_m$ ($m \ge 3$) such that $\b(T_i) = \b_i$, $i=1\ldots m$ with the two extra conditions $\b'(T_1) = \b'(T_m)=0$.  The reason for the extra conditions is that $T_1$ has always been chosen as the first moment in which cases in Italy were reported; it seems then reasonable to assume that, before that, society was in normal conditions and contact rate was constant. Also near the end of the current period of analysis, society seems to have adjusted to current `lock-down' conditions, so that contact rate is constant. Clearly, analysis of different period could suggest other boundary conditions, for instance natural splines.

Since \eqref{Volterra} is a time-dependent model, we cannot properly speak of the reproduction number $R_0$ for that model. However, we can define $R_0(t)$ as the value of $R_0$ for the model, assuming that the contact rate were frozen to the value $\b(t)$. Since in all the examples $\phi(\cdot)$ is the density of a probability distribution, we obtain $R_0(t) = \b(t)$.

Parameters were estimated by using maximum likelihood on \eqref{likelihood}, using the package \texttt{bbmle} of R \cite{bbmle}. Confidence intervals for all parameters were obtained with the same package using profile likelihood methods \cite{Bolker2008}. 

\section{Results}
We applied the parameter estimation method discussed above to the Italian dataset. We chose as reference final day of fit April 6, and start of the simulations on February 1, since it has been establihed that around that day new infections started to occur continuously \cite{Cereda2020}. The function $j_0$ providing the initial input to the model has been chosen to be equal to 1 (remember that the fit introduces then a multiplicative constant) up to $t=2$ and to 0 from $t=3$ onwards.\\
As stated above, we estimated the function $\b(t)$ as a cubic spline with either 3, 4 or 5 knots. In all cases the first knot was set on February 22 (the day after the information was released of the first confirmed case of local transmission in Italy), and the last one on March 31 (a few days before the final day). In case there were 5 knots the intermediate dates were March 3, 13 and 23 (general lock-down was decided between March 8 and 11, depending on the localities, and strengthened on March 23); with 4 knots March 6 and 18; with 3 knots March 13.  Timings

First of all we tested whether there was a difference in data fit between the different choices of delay functions. We show in Table \ref{tab:uno} the AIC obtained with the possible combinations of functions $f_1(\cdot)$ and $f_2(\cdot)$ and generation time $\phi$ applied to the Italian data (similar results are obtained with regional data) fitting on hospitalizions and death data ($q = 1/2$).\\
 \begin{table}[htp]
\caption{AIC obtained using the Poisson likelihood and different delay functions. Spline knots were 5.}
\begin{center}
\begin{tabular}{cl|c|c|c|c|c|}
&&\multicolumn{5}{c}{Generation time}\\ 
&&\multicolumn{2}{c|}{Flaxman}&\hglue1cm&\multicolumn{2}{c|}{Cereda}\\ 
&&\multicolumn{2}{c|}{$f_1$}&\hglue1cm&\multicolumn{2}{c|}{$f_1$} \\
&&Lin+Pel&Lin+Lin&&Lin+Pel&Lin+Lin\\ \hline
&Lin-Lin&838&1018 & &838.4&1018\\ 
$f_2$&&&&&&\\
&Lin-Flax&1208.4&1415.8&&1206.4&1415.8 \\
\end{tabular}
\end{center}
\label{tab:uno}
\end{table}%
 It is apparent that the fit with the function $f_1$ `Linton+Pellis' and $f_2$ `Linton+Linton" is much better than with all alternatives; hence, only that one will be considered in what follows. On the other hand, the model fit does not allow to discriminate the generation time distribution (also parameter estimates are similar). For the sake of simplicity, we will only show the results obtained with the function $\phi$ proposed by Flaxman \textit{et al.} \cite{Flaxman2020}.

Most results have been obtained applying the estimation to the data of the whole Italy and of three regions of Northern Italy (Lombardy, Emilia-Romagna and Veneto). The first two are the regions with the most confirmed cases and deaths; Veneto is currently fourth among regions as for number of confirmed cases and deaths, but was the second region where a case of local transmission was identified; moreover, according to several reports, the response to the emergency was prompt and effective, which shows also in the quality of reported aggregated data.

\begin{figure}[hp]
\begin{center}
\begin{tabular}{cc}
\includegraphics[width=6cm]{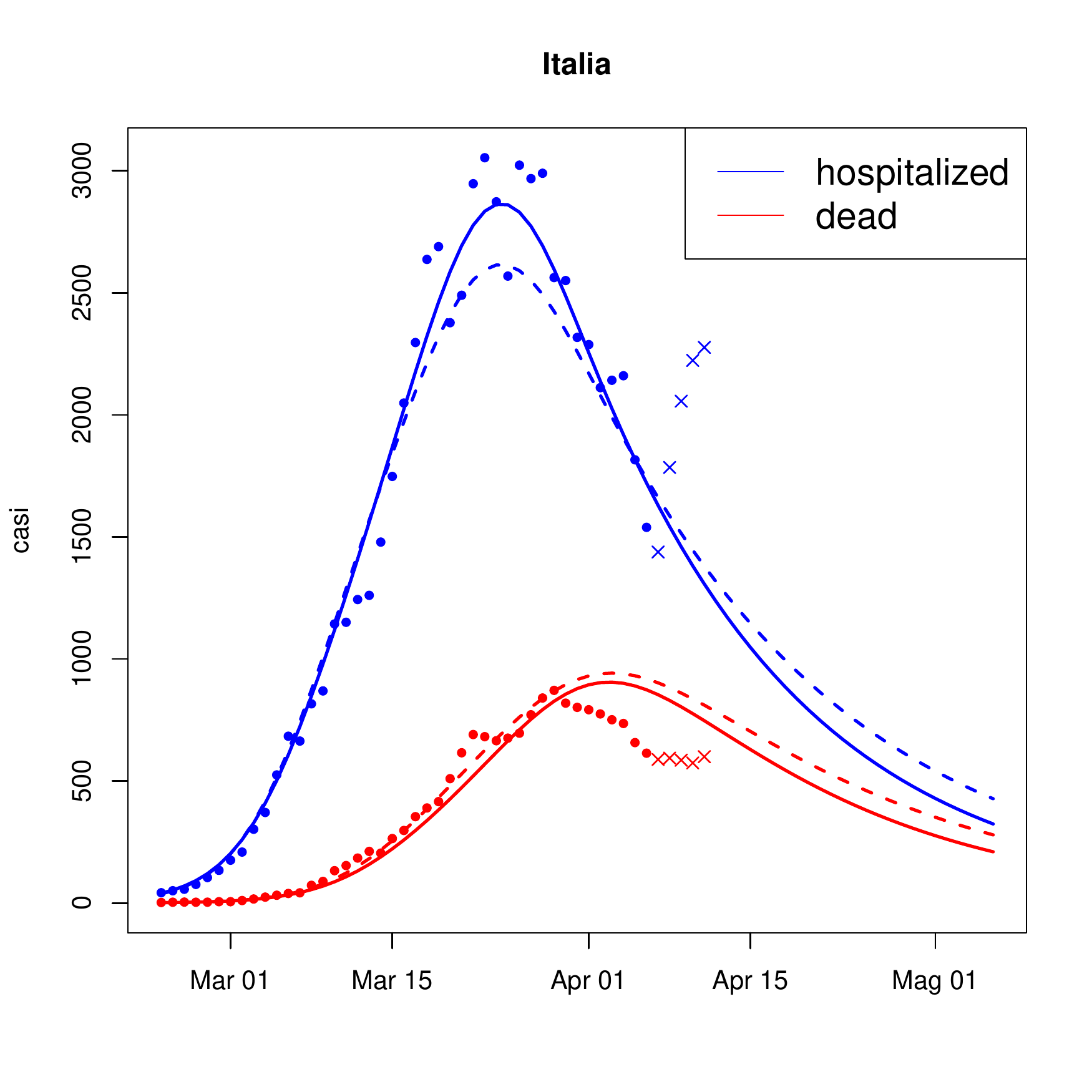}& 
\includegraphics[width=6cm]{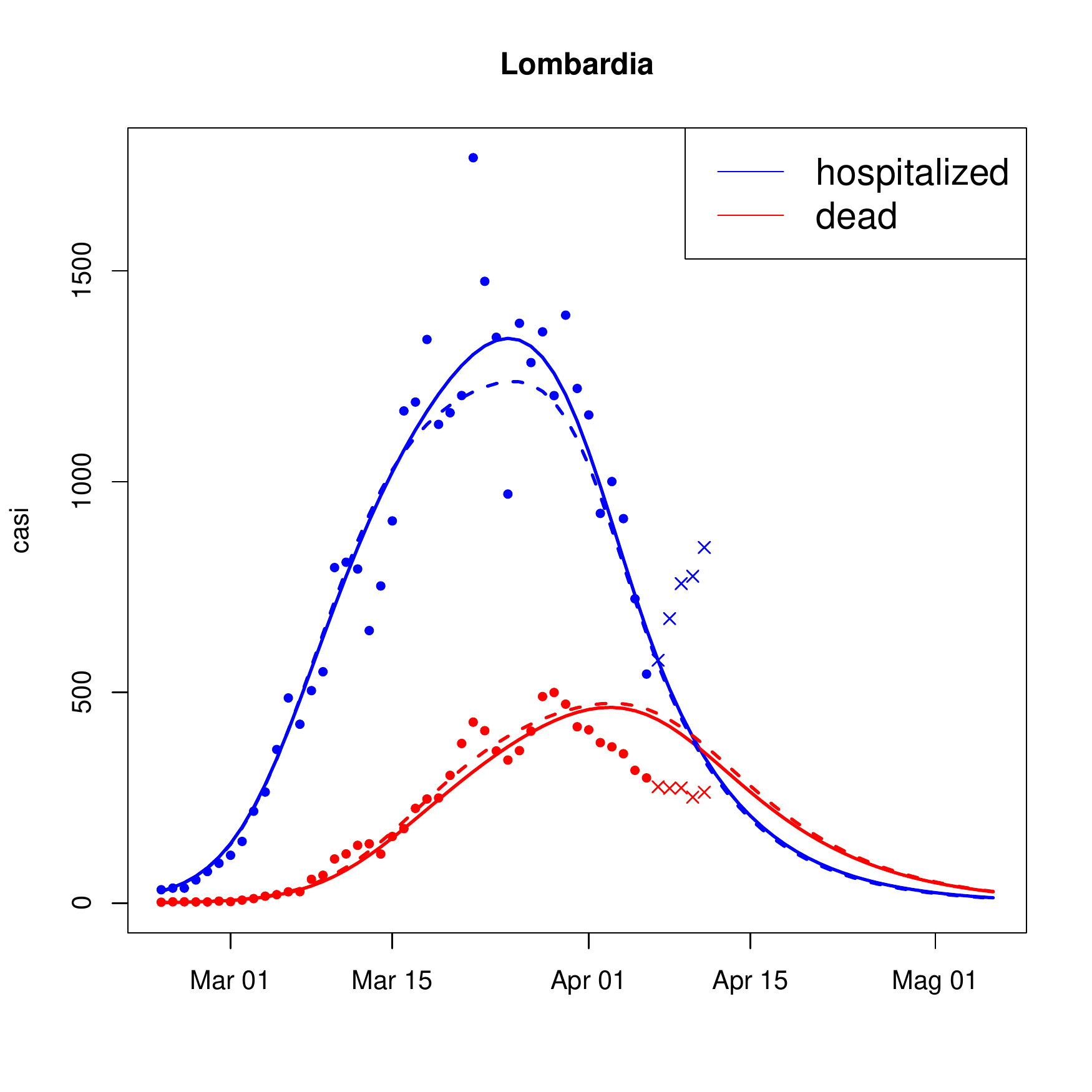}\\
\includegraphics[width=6cm]{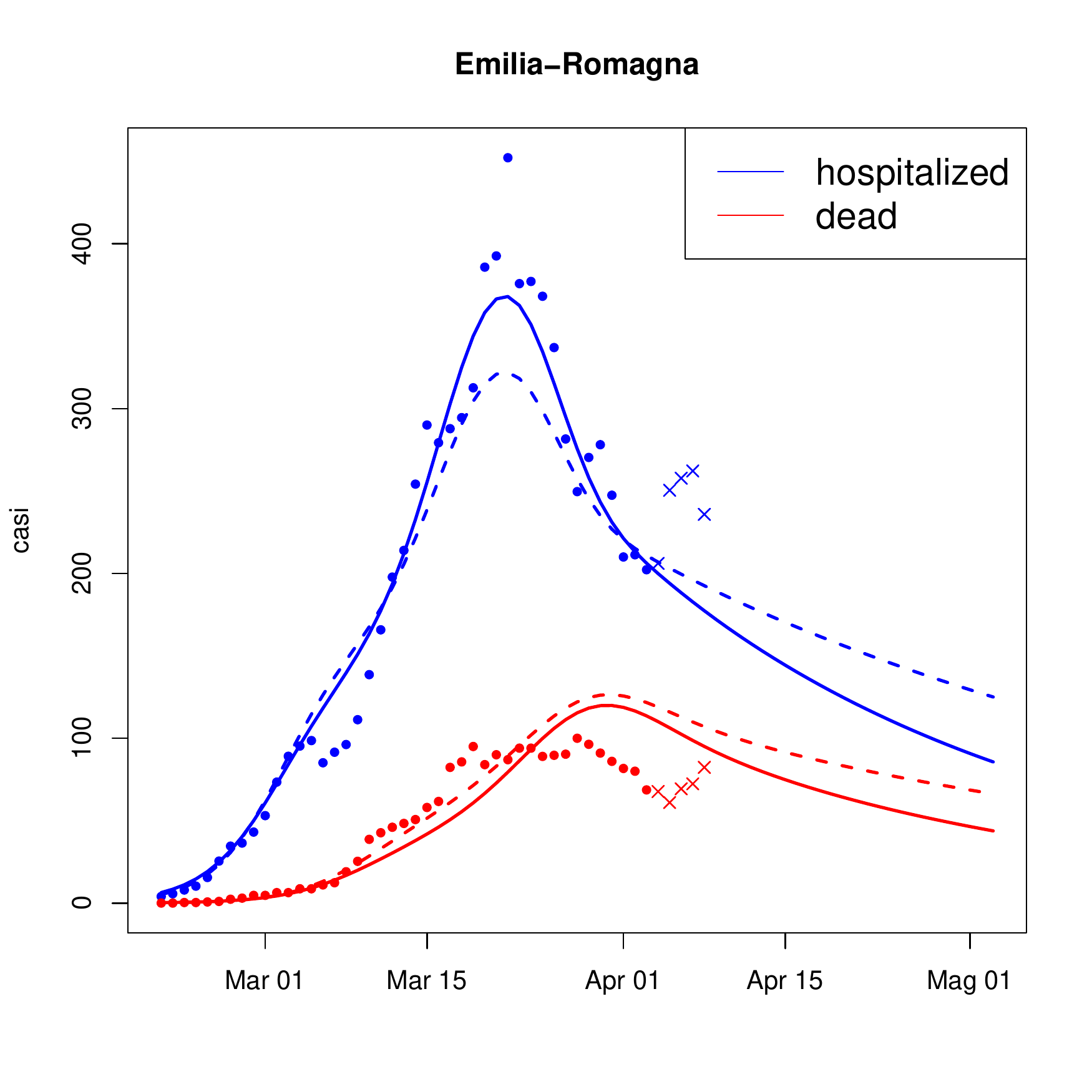}& 
\includegraphics[width=6cm]{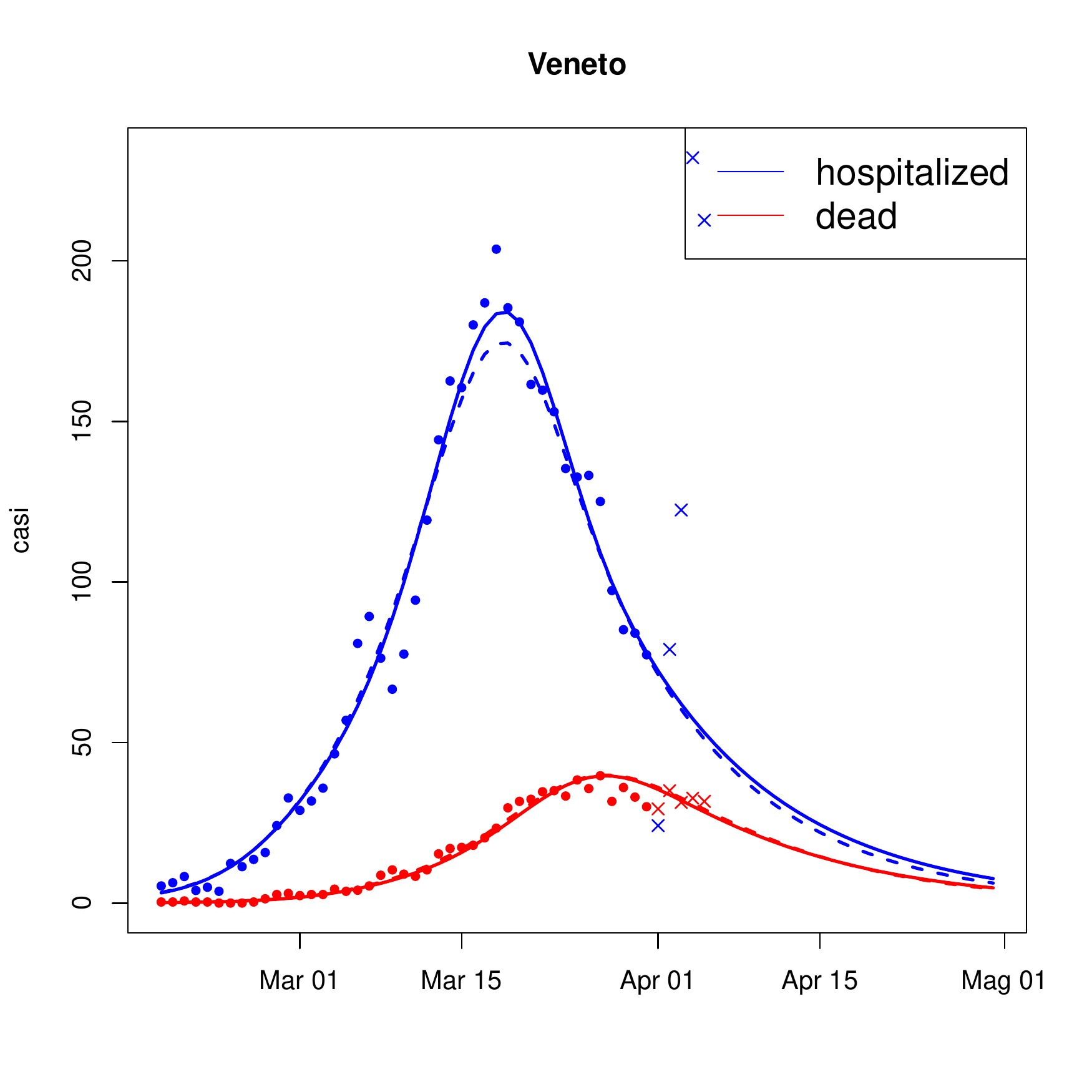}\\
\end{tabular}
\caption{The points represent hospitalizations (blue) and deaths (red)  from the whole Italy and three regions. The solid curves are the corresponding fitted models using the Poisson likelihood; the dashed curves the corresponding fitted models using the negative binomial likelihood (with $k=3$). Dots correspond to data (up to the final day) used in the fit; crosses data beyond the final day, not used in fitting} 
\label{fig:ref_fit}
\end{center}
\end{figure}

\begin{figure}[H]
\begin{center}
\begin{tabular}{cc}
\includegraphics[width=6cm]{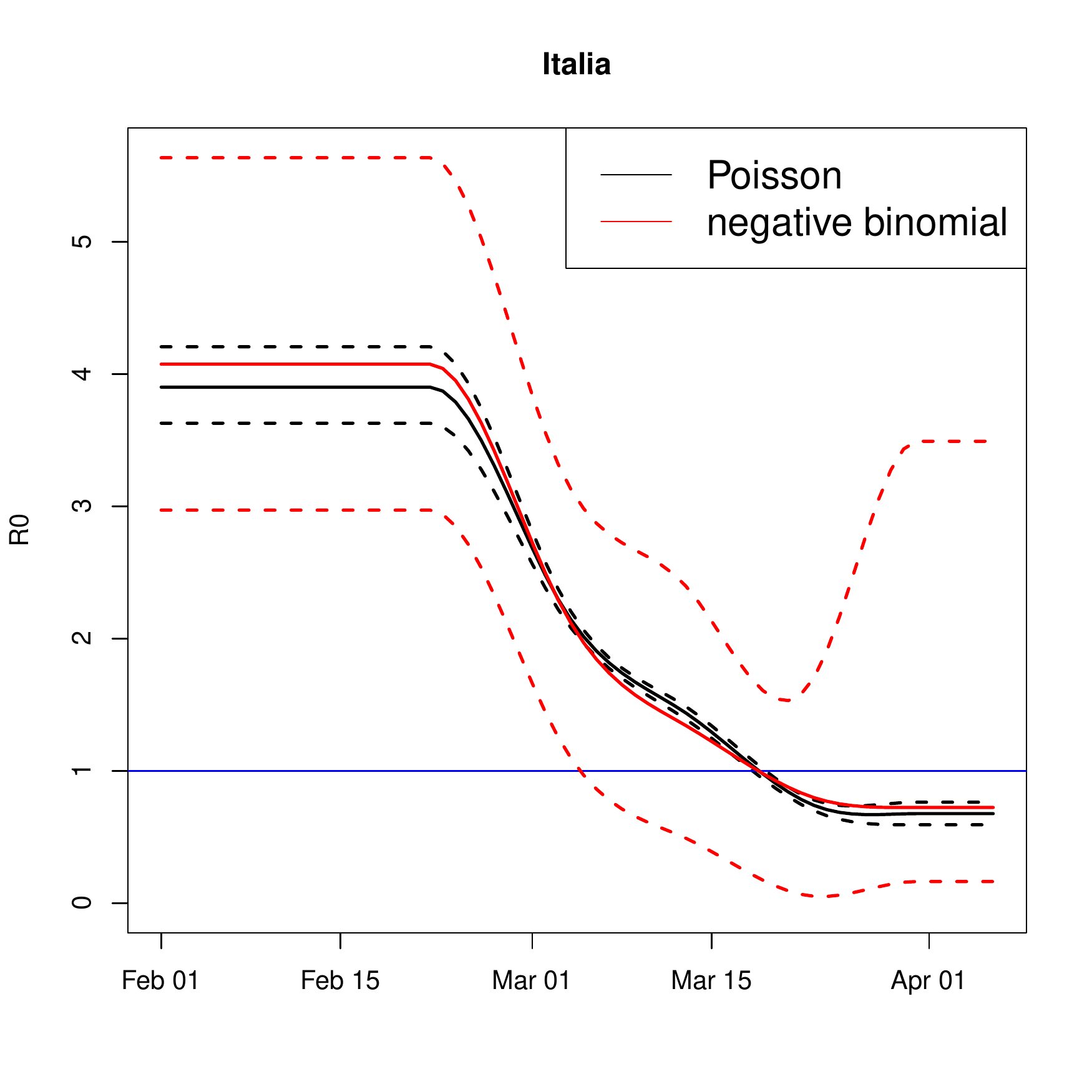}& 
\includegraphics[width=6cm]{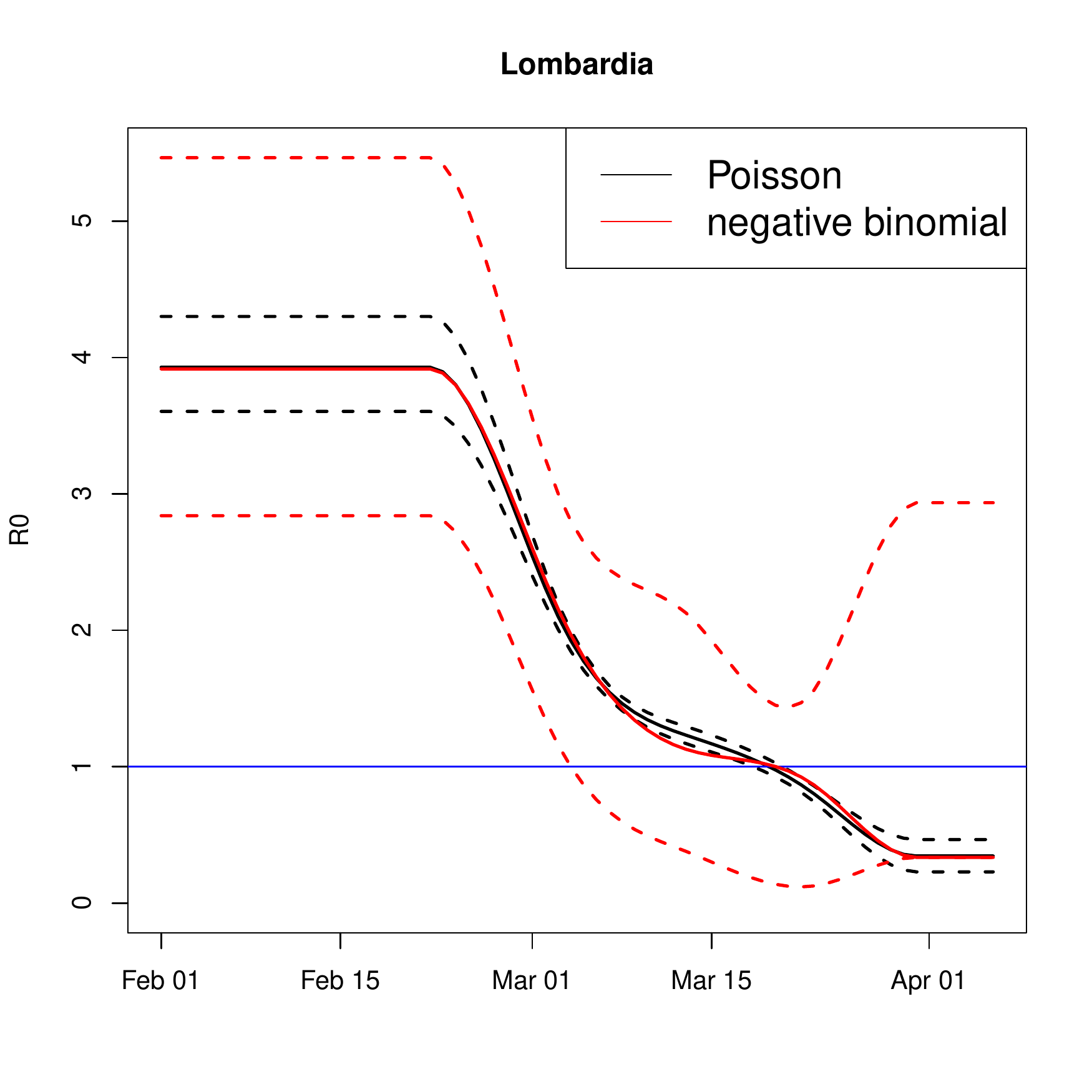}\\
\includegraphics[width=6cm]{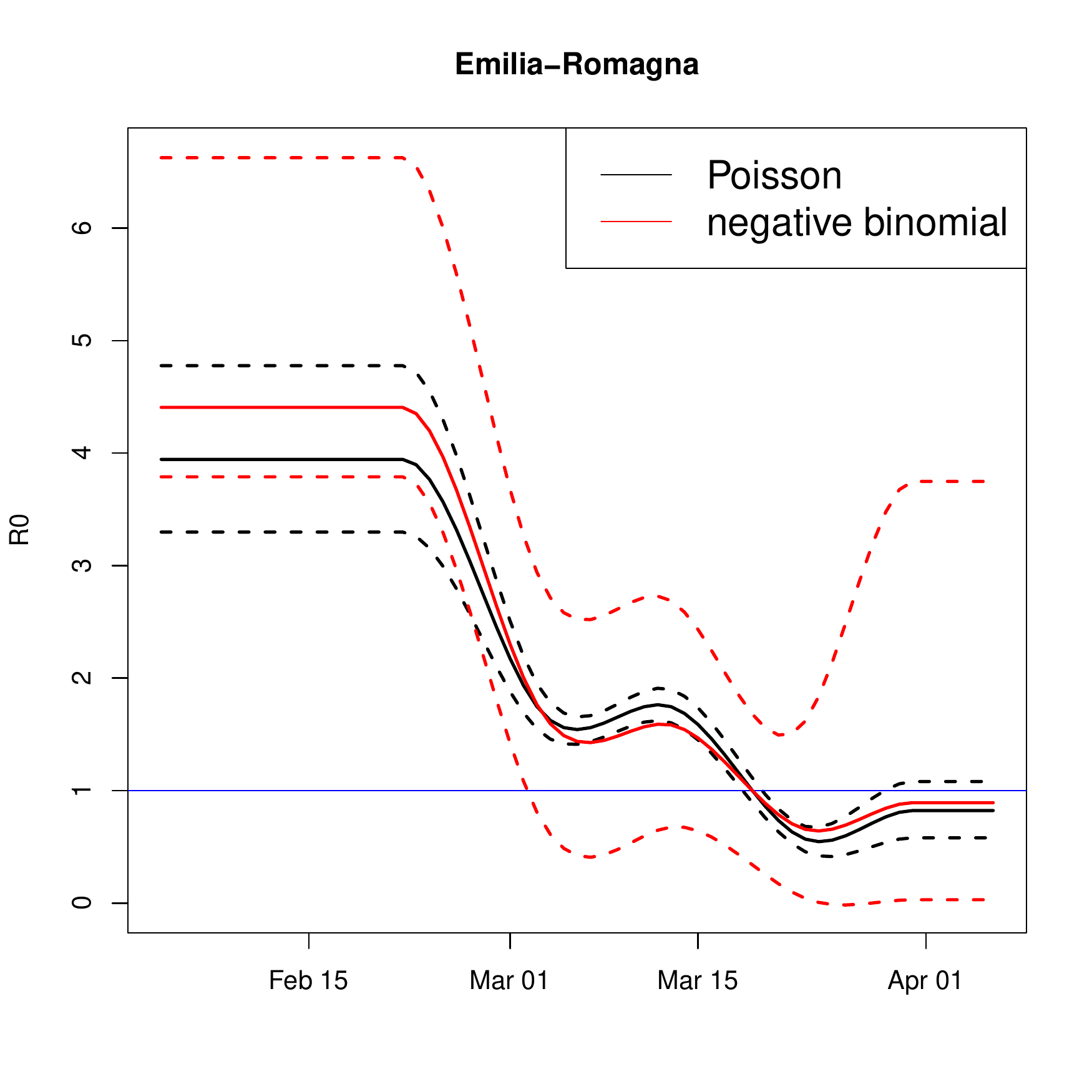}& 
\includegraphics[width=6cm]{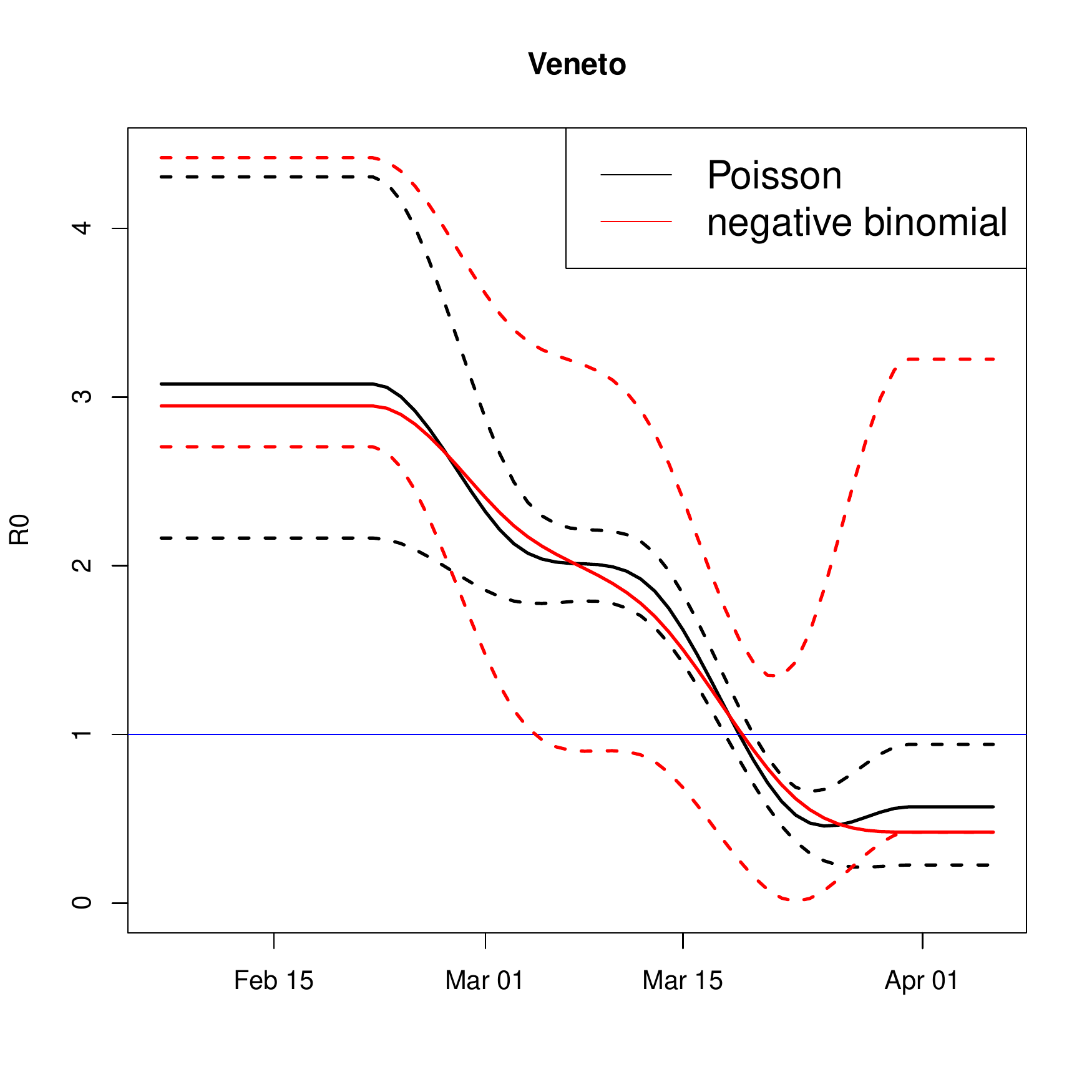}\\
\end{tabular}
\caption{Estimated values of $R_0(t)$ for the whole Italy and three regions. The solid curves represent estimated values; dashed curves corresponding confidence intervals (see text for details). The values obtained using the Poisson likelihood are shown in black; those using the negative binomial likelihood in red.} 
\label{fig:ref_R0}
\end{center}
\end{figure}

First we show the fit, using either the Poisson or negative binomial likelihood, to data of the four geographic units (Fig.~\ref{fig:ref_fit}) and the resulting estimates of $R_0$ (Fig.~\ref{fig:ref_R0}) with the corresponding confidence intervals. To be precise, the confidence intervals refer only to the values in the knots (Table \ref{tab:2}); the curves are obtained by fitting cubic splines to the extremes of the confidence intervals, but do not properly represent confidence intervals for the times outside the knots. The negative binomial was fitted, assuming a fixed value $k=3$ of the dispersion parameter, value chosen after some trials.

\begin{table}[htp]
\caption{Values of $R_0$ at the 5 knots (CI from Poisson likelihood)}
\begin{center}
\begin{tabular}{cc}
Italy & Lombardia \\ \hline
\begin{tabular}{l c|c}
& estimate & 95\% CI \\
February 22&	3.901&	3.628 - 4.206\\
March 3&2.3	&2.222 -	2.377\\
March 13&1.437&	1.389 -	1.485\\
March 23&0.739&	0.698 -	0.779\\
March 31&0.676&	0.592	- 0.763
\end{tabular}
& \begin{tabular}{c|c}
 estimate & 95\% CI \\
	3.928&	3.604 -		4.301\\
2.103	&2.008 -		2.197\\
1.231&	1.169 -		1.295\\
0.801&	0.743 -		0.859\\
0.345&	0.229 -		0.466
\end{tabular}
\end{tabular}\\[1em]
\begin{tabular}{cc}
 Emilia-Romagna & Veneto\\ \hline
\begin{tabular}{l c|c}
& estimate & 95\% CI \\
February 22&	3.944&	3.298 -		4.778\\
March 3&1.746	&1.548 -		1.948\\
March 13&1.746&	1.602 -		1.898\\
March 23&0.568&	0.456 -		0.681\\
March 31&0.823&	0.58 -		1.081
\end{tabular}
& \begin{tabular}{c|c}
 estimate & 95\% CI \\
	3.078&	2.164 -		4.305\\
2.131	&1.79 -		2.497\\
1.849&	1.634 -		2.075\\
0.521&	0.363 -		0.686\\
0.57&	0.226 -		0.941
\end{tabular}
\end{tabular}\\

\end{center}
\label{tab:2}
\end{table}%
One can see from Figures \ref{fig:ref_fit} and \ref{fig:ref_R0} that the central estimates are extremely similar for the two likelihood; the main difference is in the confidence intervals that are much wider for the negative binomial likelihood, as expected. From the AIC one should choose the Poisson model for the Veneto region, and the negative binomial for the other units (although for Emilia-Romagna the difference in AIC between the two models is not large). For the sake of simplicity, in what follows we will only consider the Poisson likelihood, keeping in mind that the confidence intervals obtained are in any case only indicative, since many statistical assumptions behind the computations are not really satisfied.
\begin{figure}[htbp]
\begin{center}
\includegraphics[width=7cm]{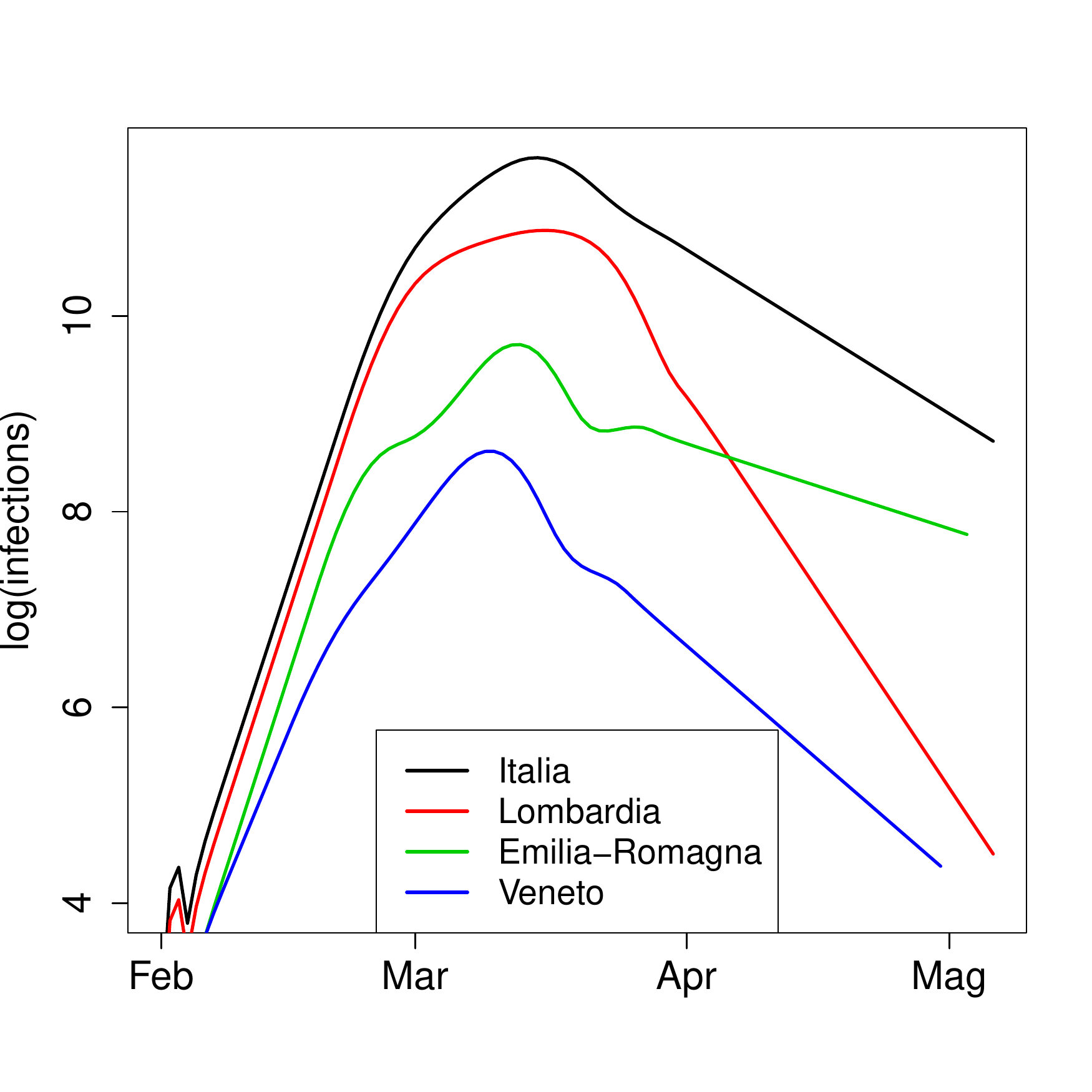}
\caption{Natural logarithm of the predicted values of new infection. The predictions are up to a multiplicative constant (an additive constant in logarithmic scale).}
\label{fig:inf}
\end{center}
\end{figure}

In Figure \ref{fig:inf} we show (in logarithmic scale) the values predicted for new infections in Italy and the three regions from February up to the beginning of May; all show a peak in new infection around March 15-20, with somewhat different patterns. Note that all the values beyond the first days of April are pure extrapolations with no statistical value.
\begin{figure}[htb]
\begin{center}
\begin{tabular}{cc}
\includegraphics[width=6cm]{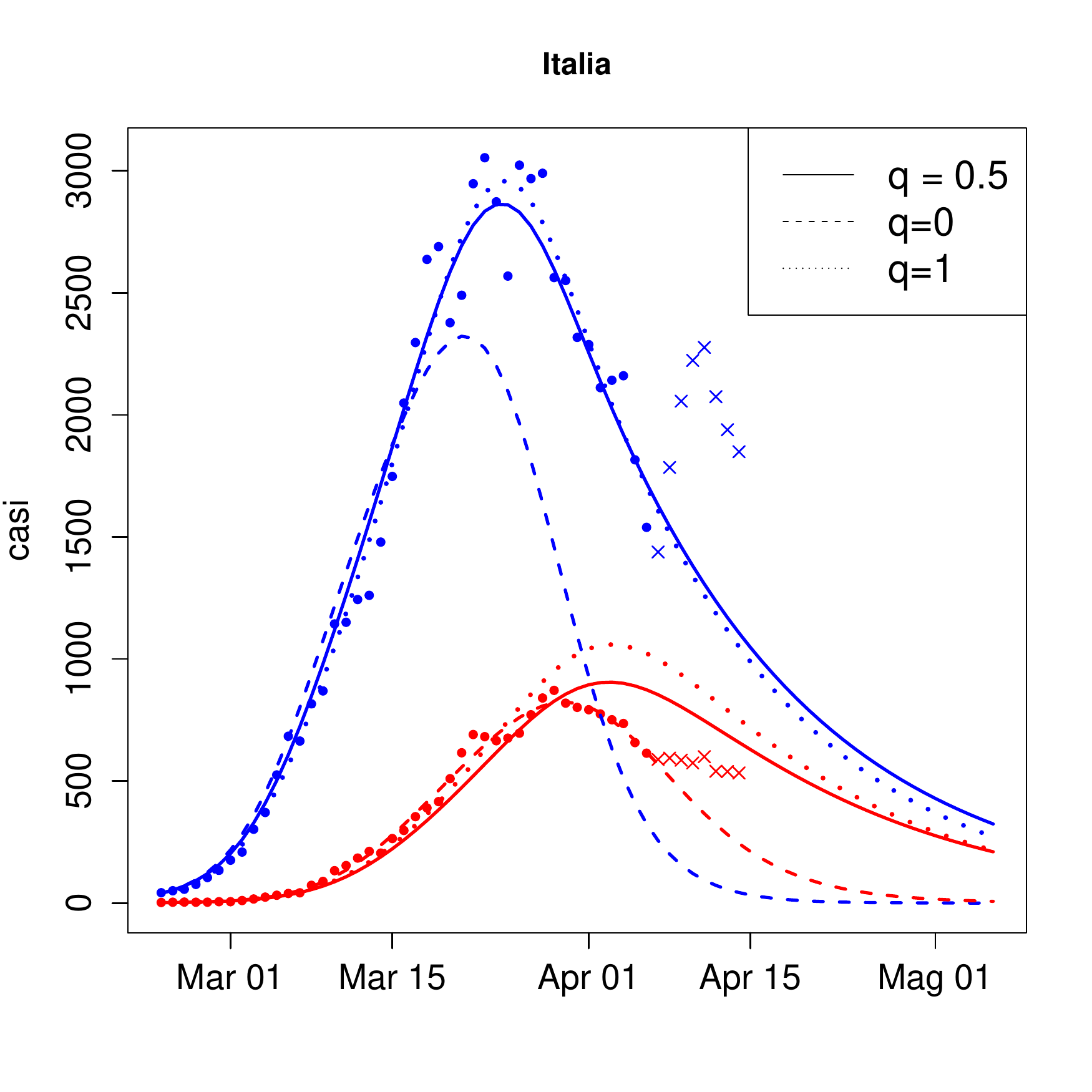}& 
\includegraphics[width=6cm]{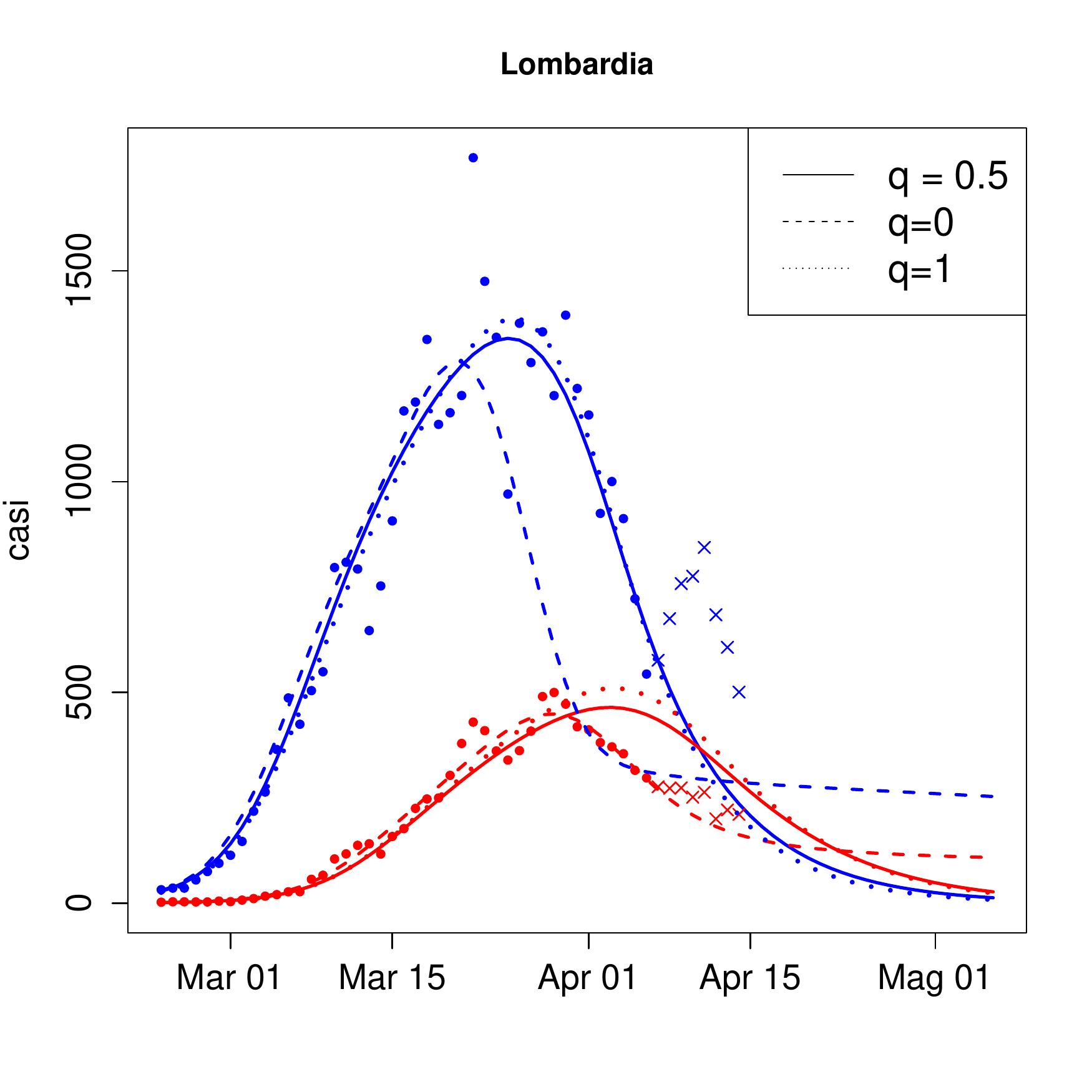}\\
\includegraphics[width=6cm]{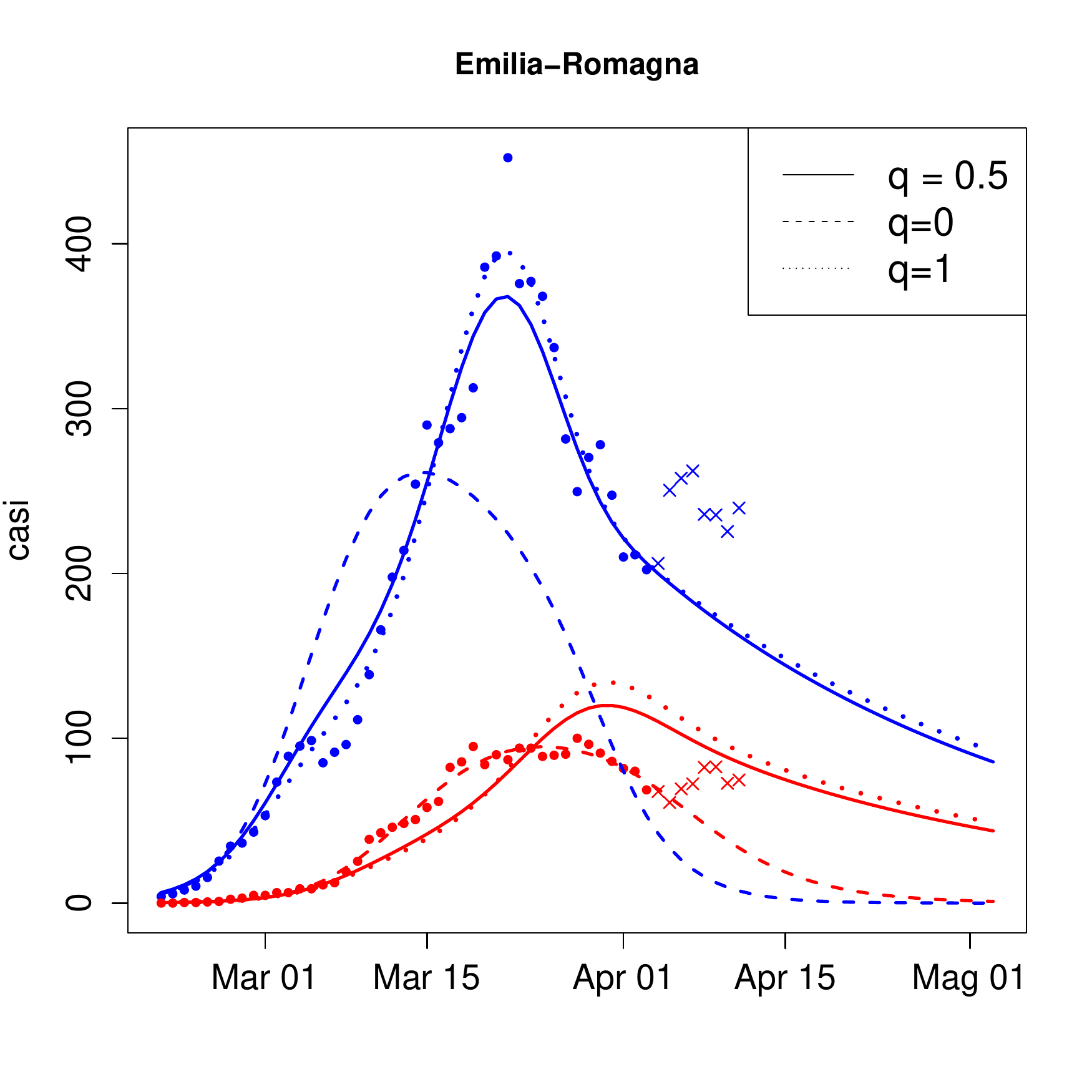}& 
\includegraphics[width=6cm]{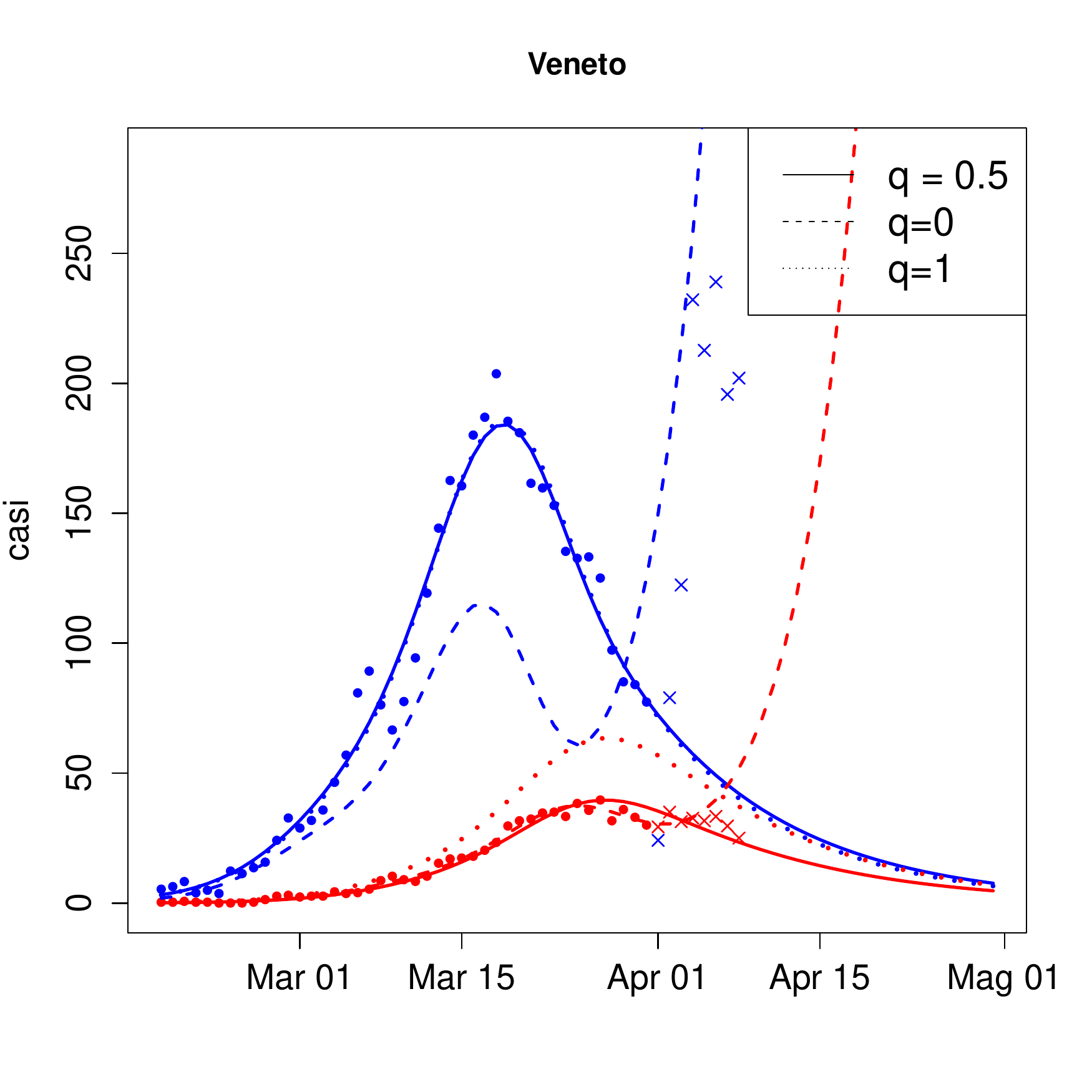}\\
\end{tabular}
\caption{The points represent hospitalizations (blue) and deaths (red)  from the whole Italy and three regions. The curves are the fitted models with $q=1/2$ (solid lines), $q=0$ (dashed lines), $q=1$ (dotted lines), blue for hospitalizations, red for deaths.} 
\label{fig:weigh_fit}
\end{center}
\end{figure}

\begin{figure}[hbtp]
\begin{center}
\begin{tabular}{cc}
\includegraphics[width=6cm]{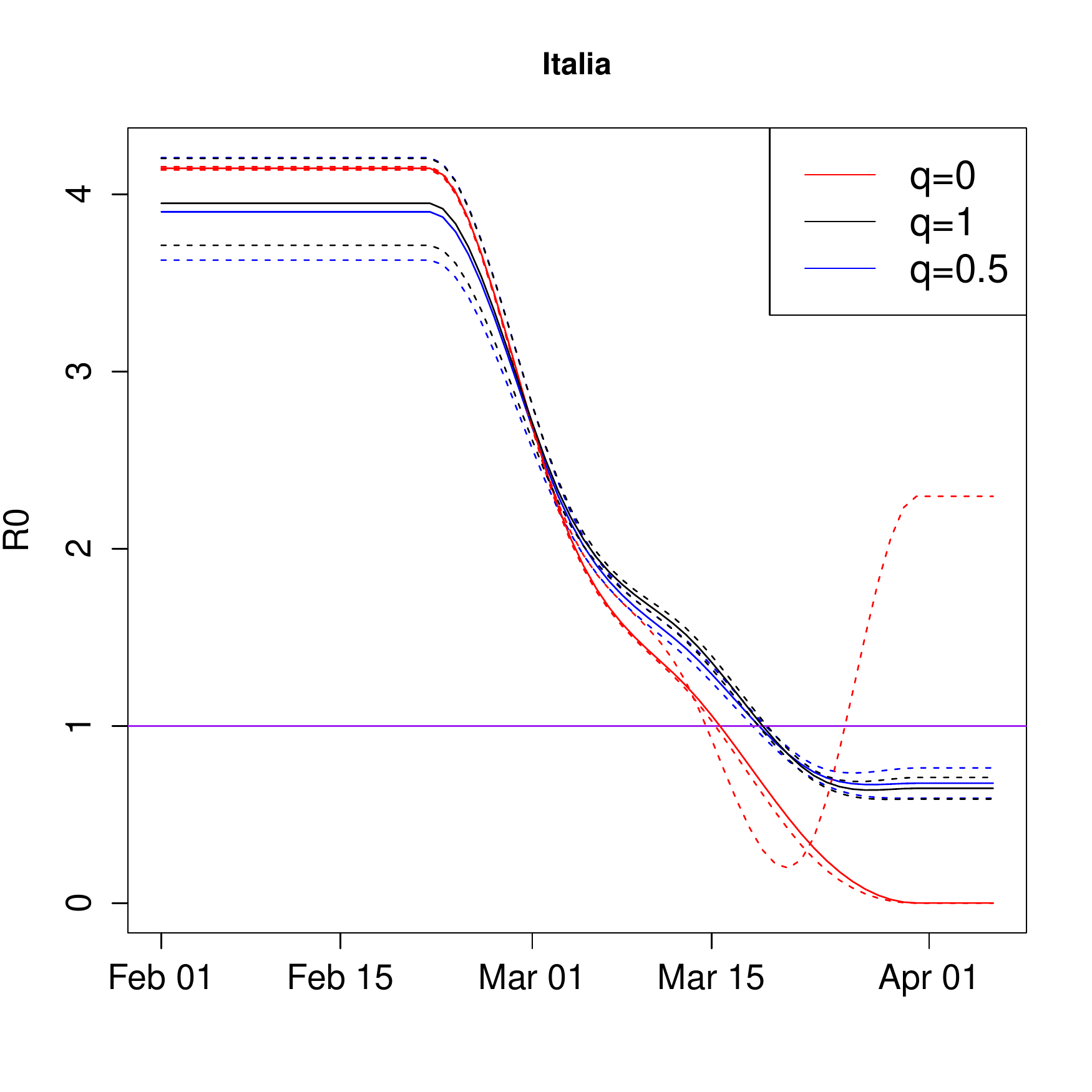}& 
\includegraphics[width=6cm]{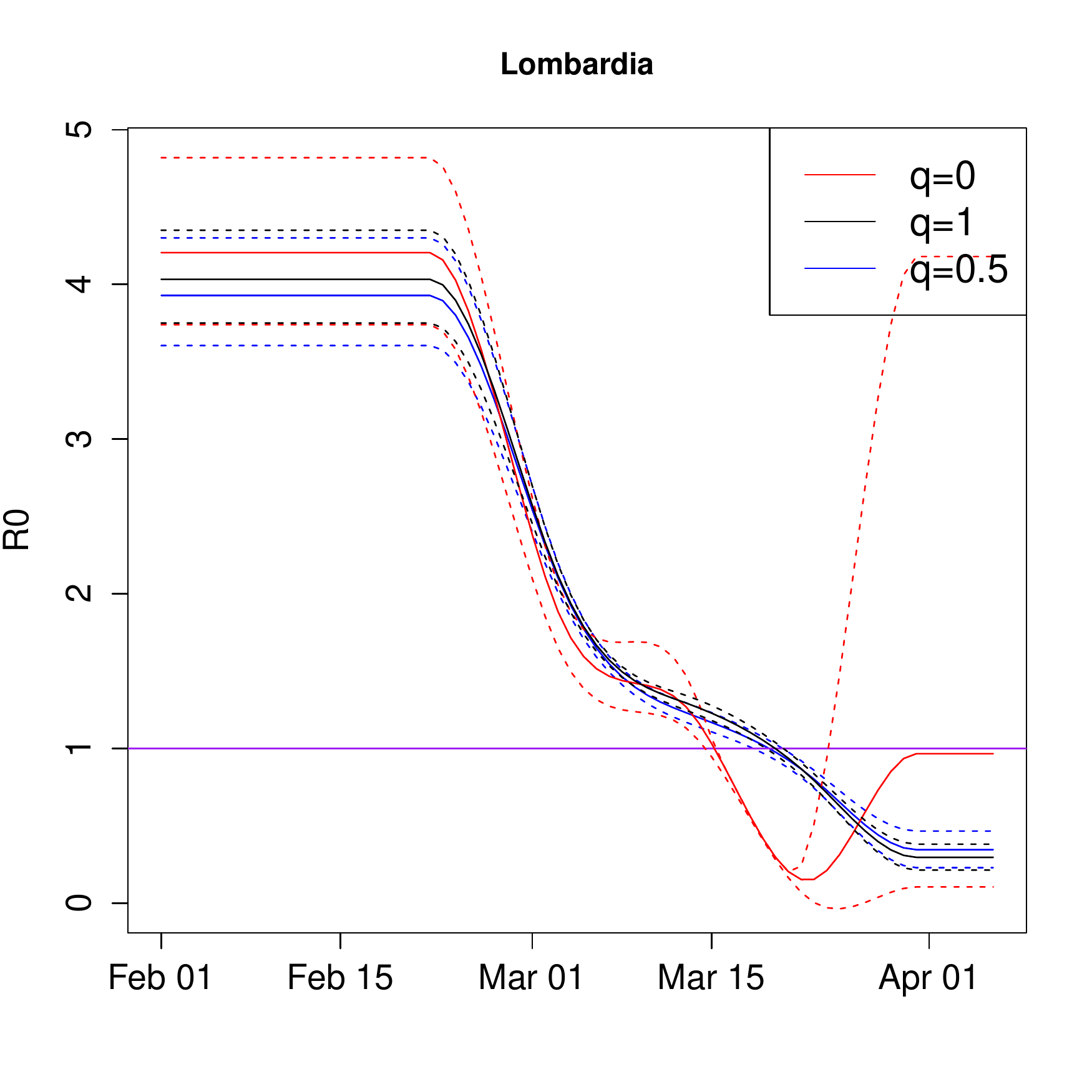}\\
\includegraphics[width=6cm]{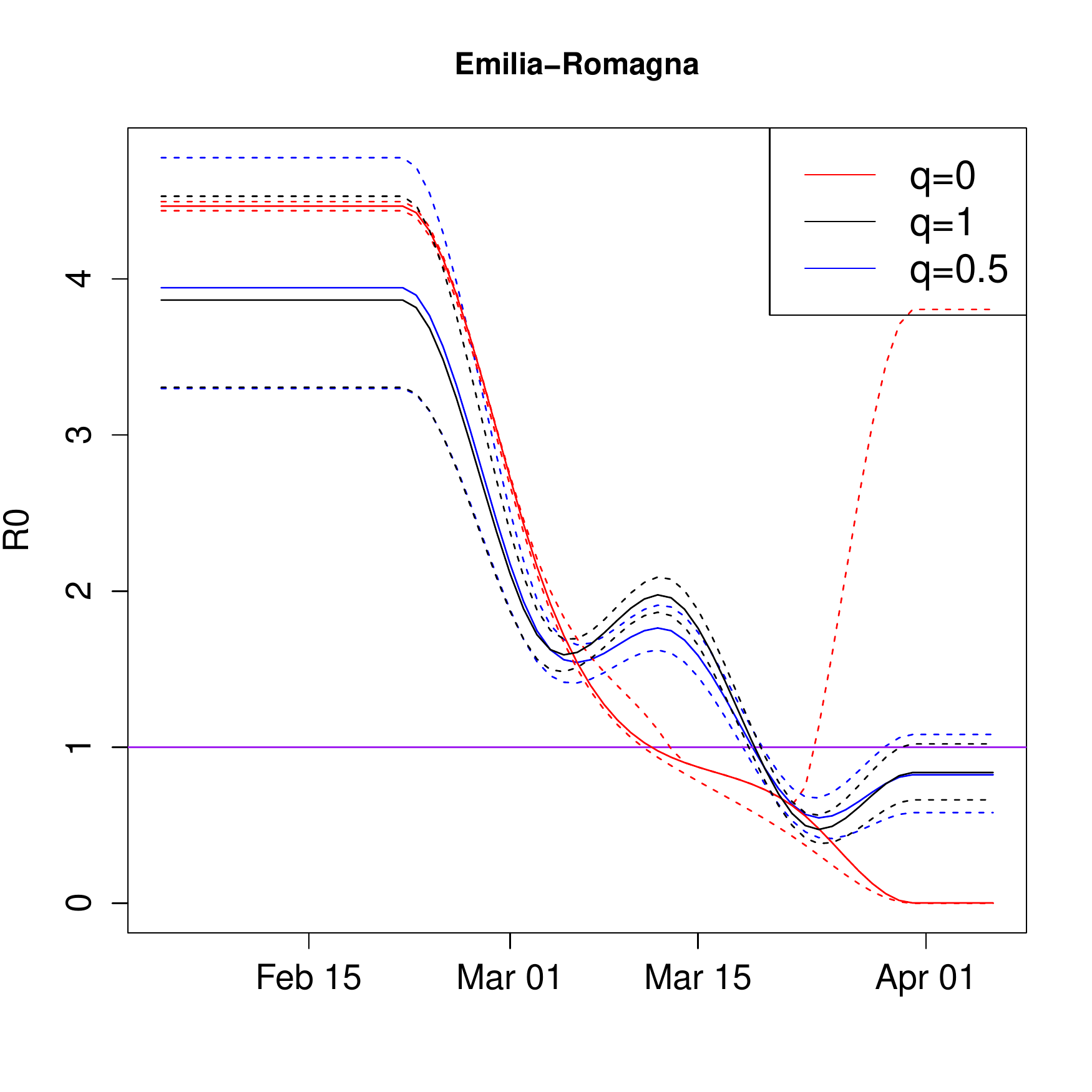}& 
\includegraphics[width=6cm]{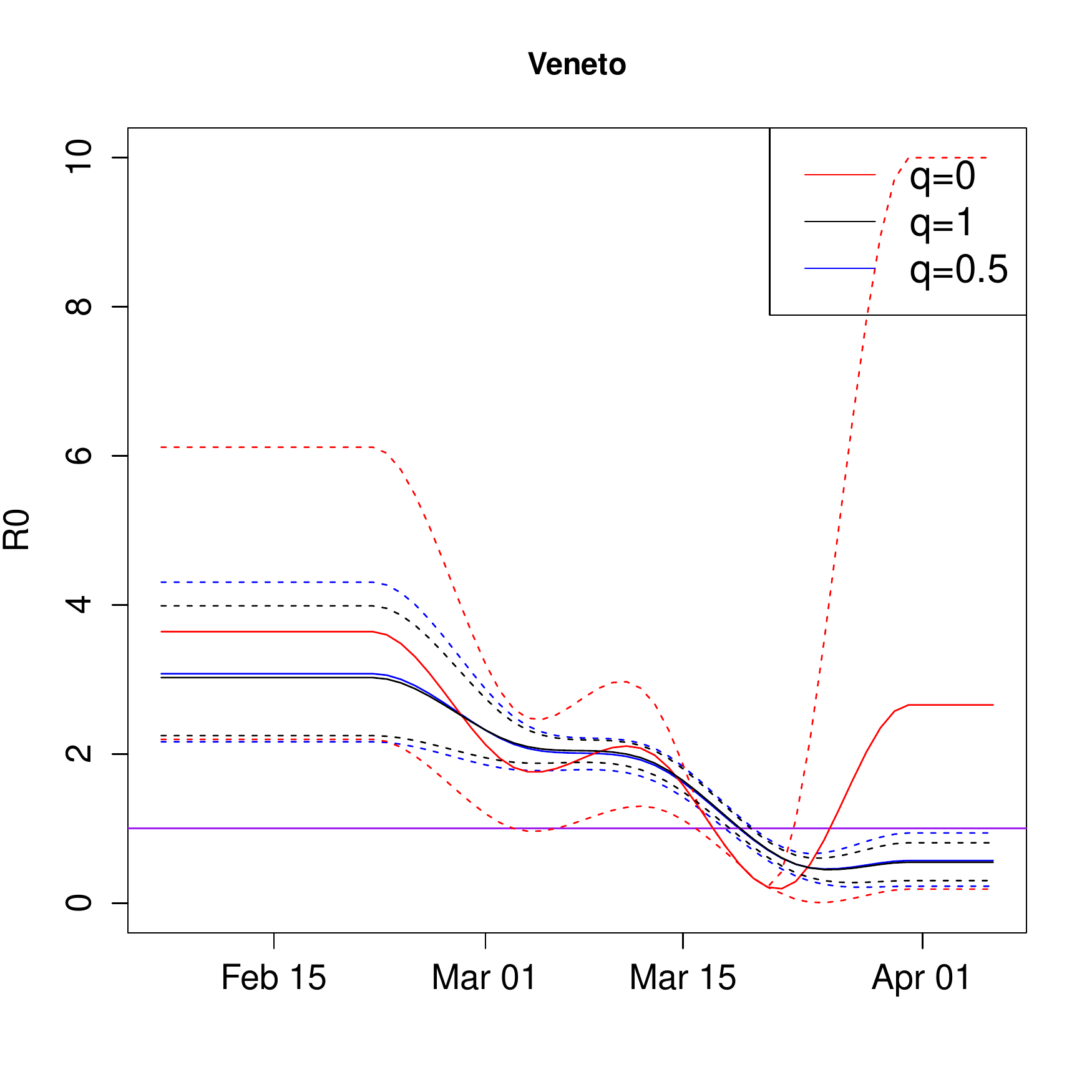}\\
\end{tabular}
\caption{Estimated values of $R_0(t)$ for the whole Italy and three regions. The solid curves represent estimated values; dashed curves corresponding confidence intervals (see text for details). Colours represent $q=0$ (red), $q=0.5$ (blue), $q=1$ (black).} 
\label{fig:weigh_R0}
\end{center}
\end{figure}

A second question asked is whether using data on both hospitalizations and deaths improves the estimates compared to using a single dataset. We show the expected values of hospitalizations and deaths in Figure \ref{fig:weigh_fit} and the corresponding values of $R_0$ in \ref{fig:weigh_R0}  for $q=0$ (fit only on deaths), $q=1$ (fit only on hospitalizations), $q=1/2$ (complete likelihood). Note that when $q=0$, predicted values of hospitalizations are irrelevant to the likelihood and are predicted only up to a multiplicative constant; the program adjusts the constant to a reasonable value, but does not attempt to fit those data; conversely, when $q=1$, for death data.

Two conclusions emerge from this analysis: first of all, the results with $q=1/2$ and $q=1$ are extremely similar, as the likelihood \eqref{likelihood} is dominated (even for $q=1/2$) from the larger numbers of hospitalizations. The other observation is that prediction using only data on deaths are much more unstable, especially for recent values of $R_0 (t)$; this can be seen by the large width  (despite the choice of Poisson likelihood) of confidence intervals for the last estimate of $\beta$ for all region.

\begin{figure}[htbp]
\begin{center}
\includegraphics[width=6cm]{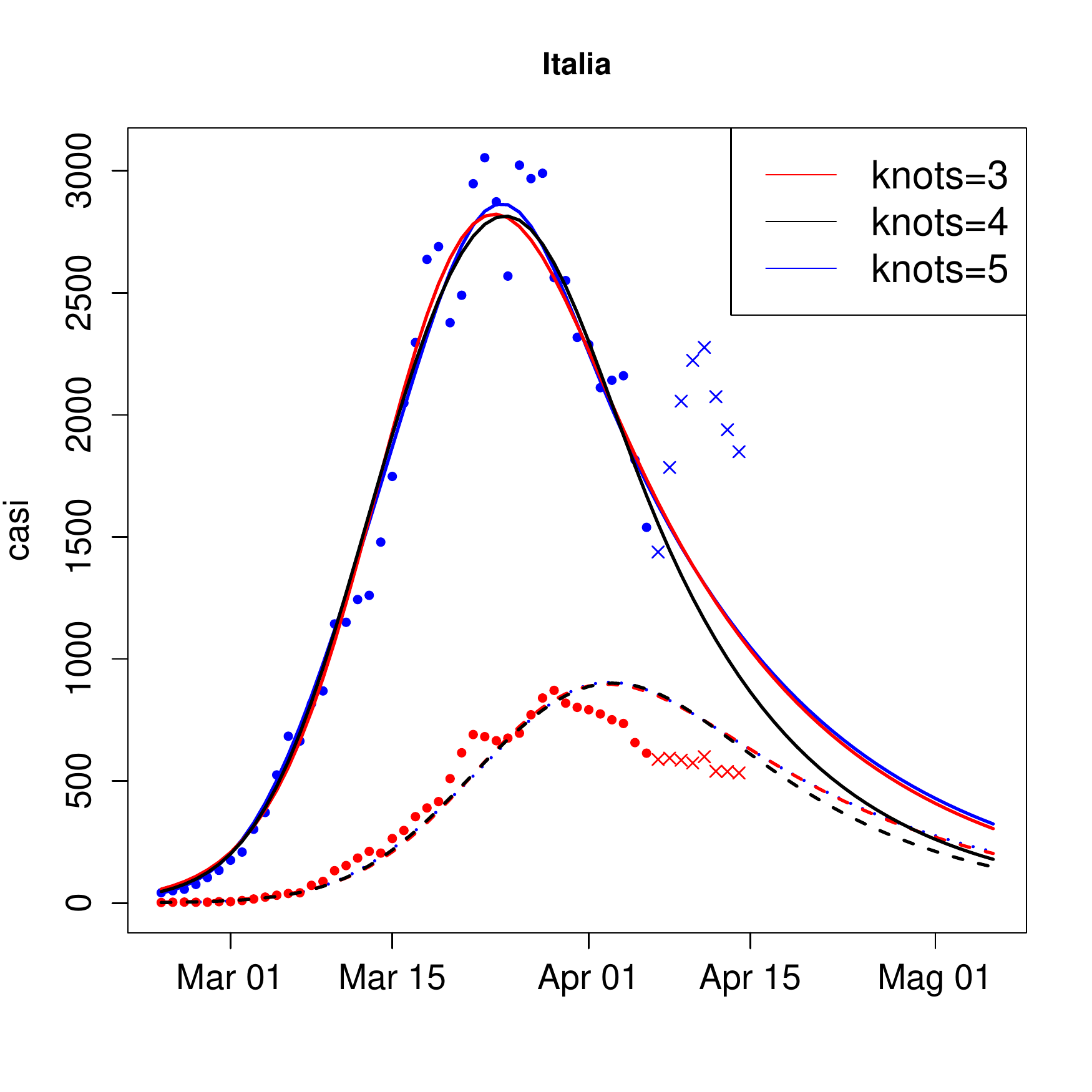}\ \includegraphics[width=6cm]{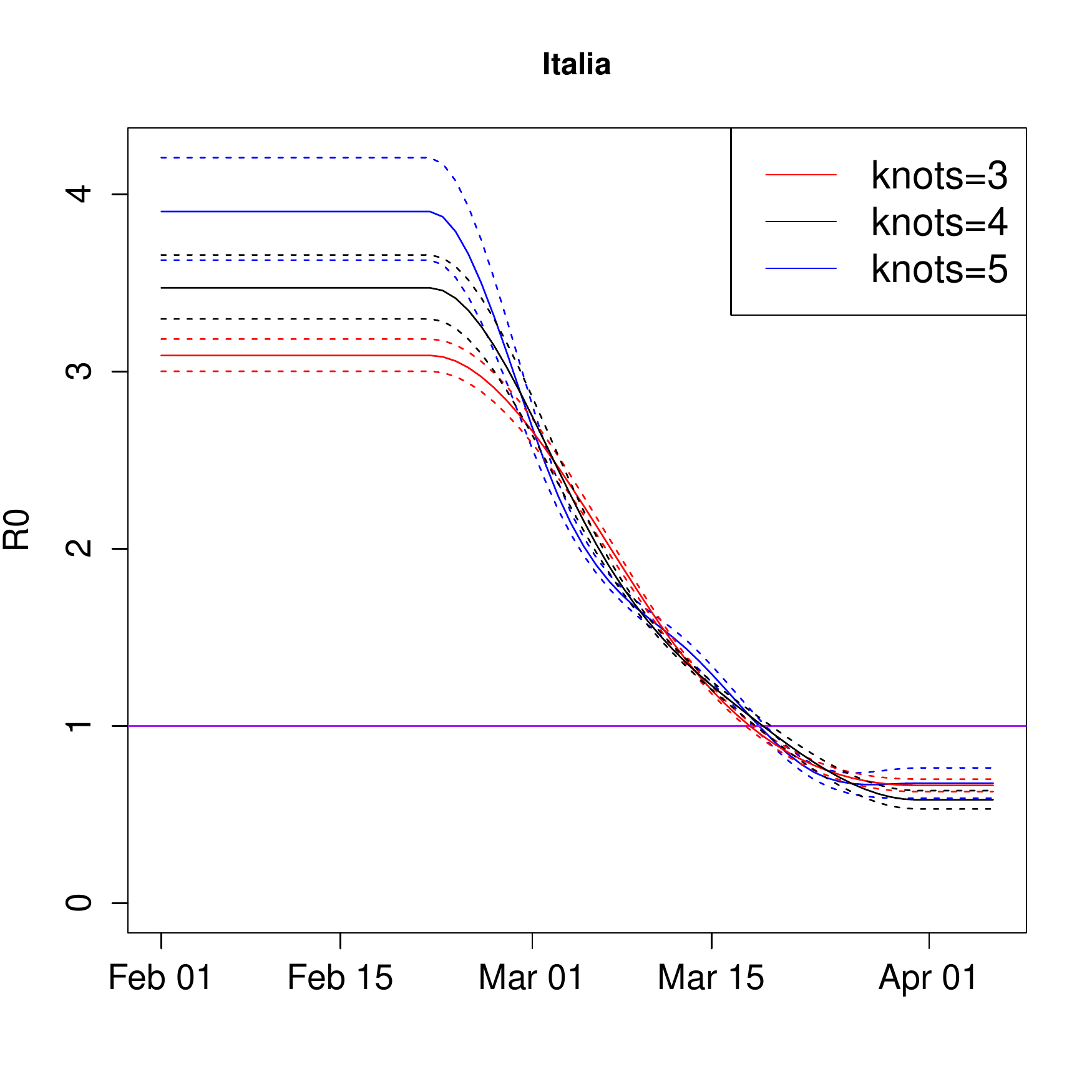}
\caption{Model fitted with 3 (red), 4 (black) or 5 (blue) knots for the splines. Left: predicted hospitalizations and deaths. Right: predicted $R_0(t)$.}
\label{fig:knots}
\end{center}
\end{figure}

\begin{figure}[htbp]
\begin{center}
\includegraphics[width=6cm]{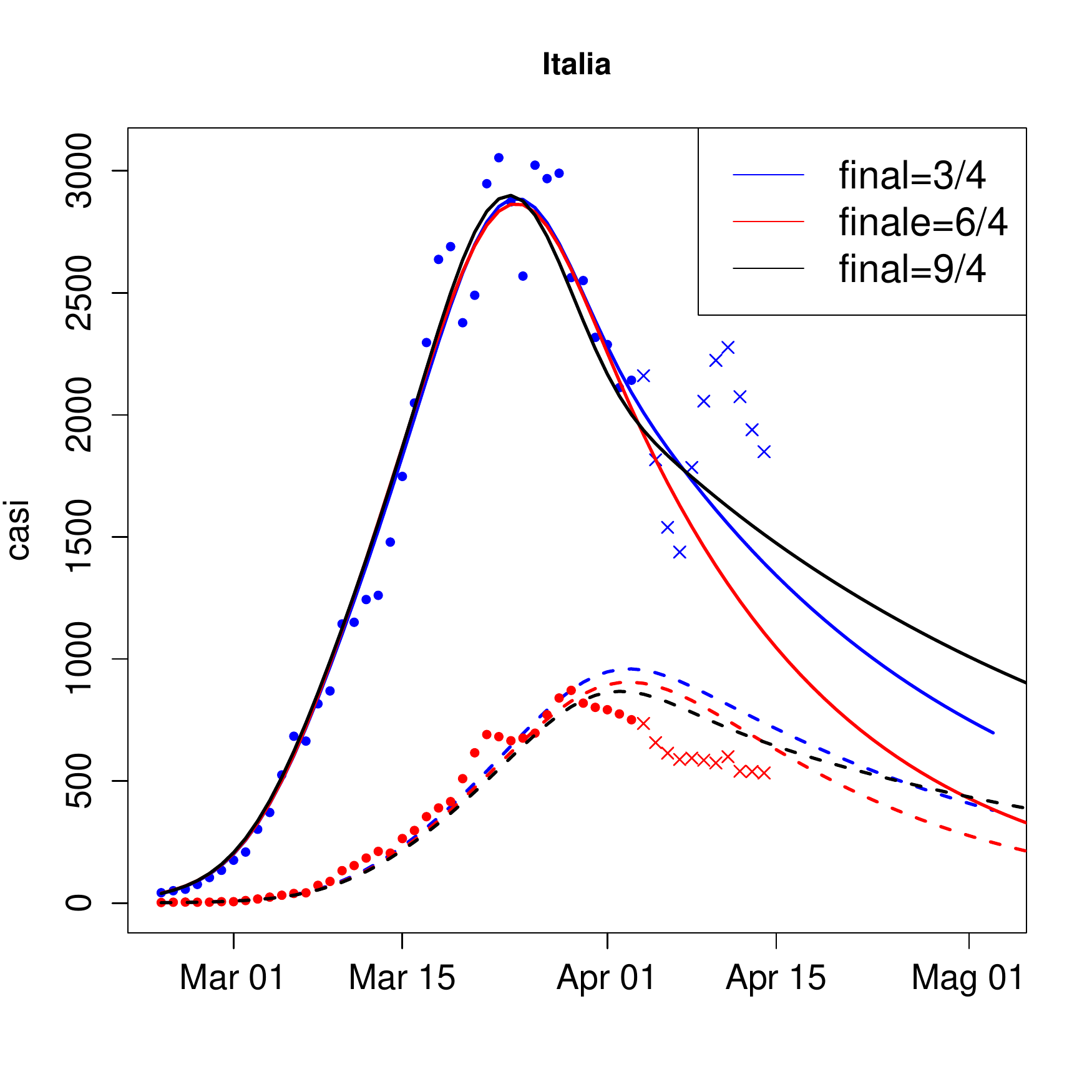}\ \includegraphics[width=6cm]{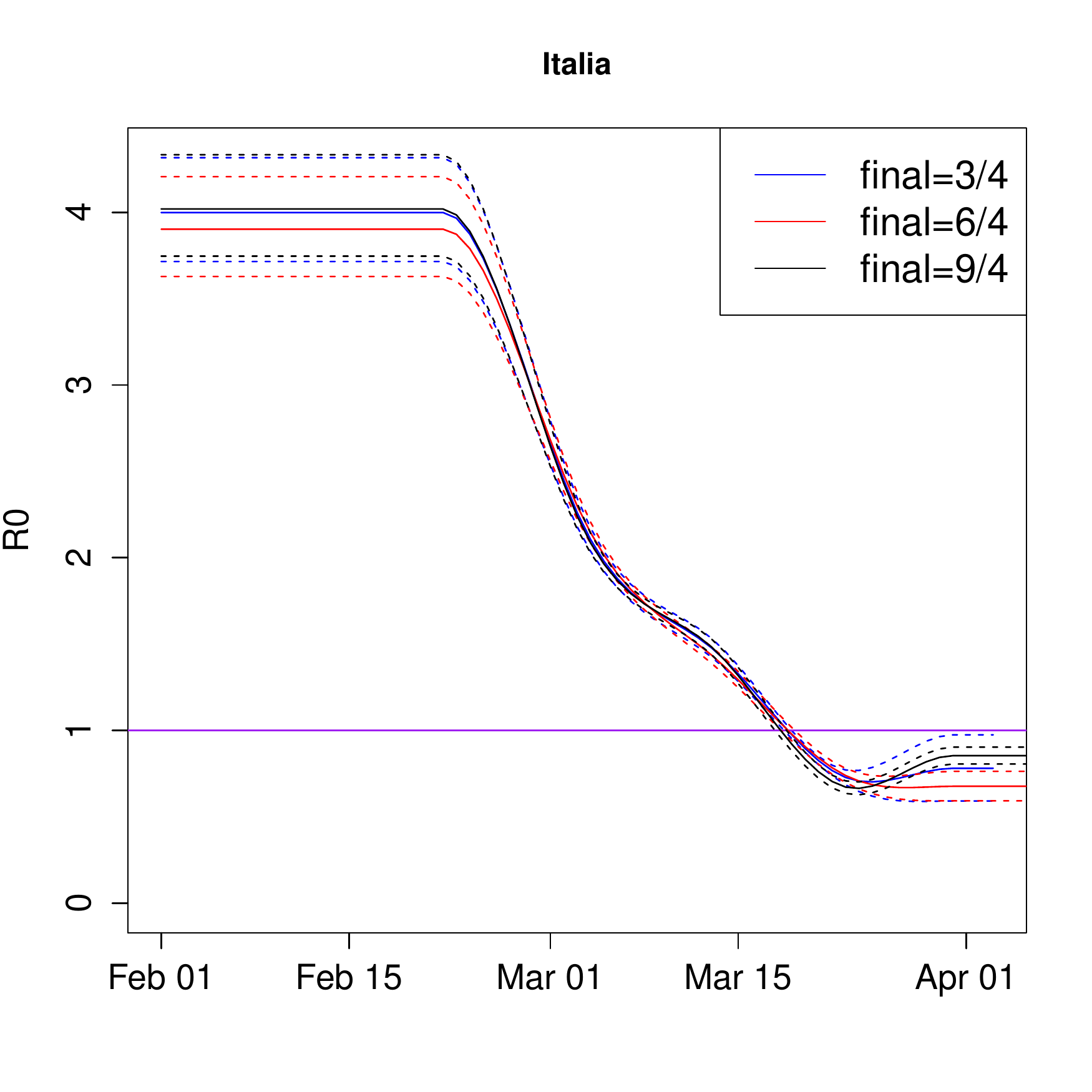}
\caption{Model predictions when fitted to data up to April 3 (blue), April 6 (red) or April 9 (black). Left: predicted hospitalizations and deaths. Right: predicted $R_0(t)$.}
\label{fig:final}
\end{center}
\end{figure}

Concentrating only on the data from the whole Italy, we finally show the effect of the number of knots (3, 4 or 5) (Fig.~\ref{fig:knots}), and of the final day of fit (April 3, 6 or 9) (Fig.~\ref{fig:final}) on the predicted values of hospitalizations, deaths and $R_0$.

In principle, the method can be used for data of any region. However, data from many regions appear very noisy, especially when absolute numbers are low that it is very difficult reasonably fitting a model to them. We only show the results obtained when applying the model to data from Piemonte and Marche (the 3rd and 5th region for number of cases).  \\
\begin{figure}[htbp]
\begin{center}
\includegraphics[width=6cm]{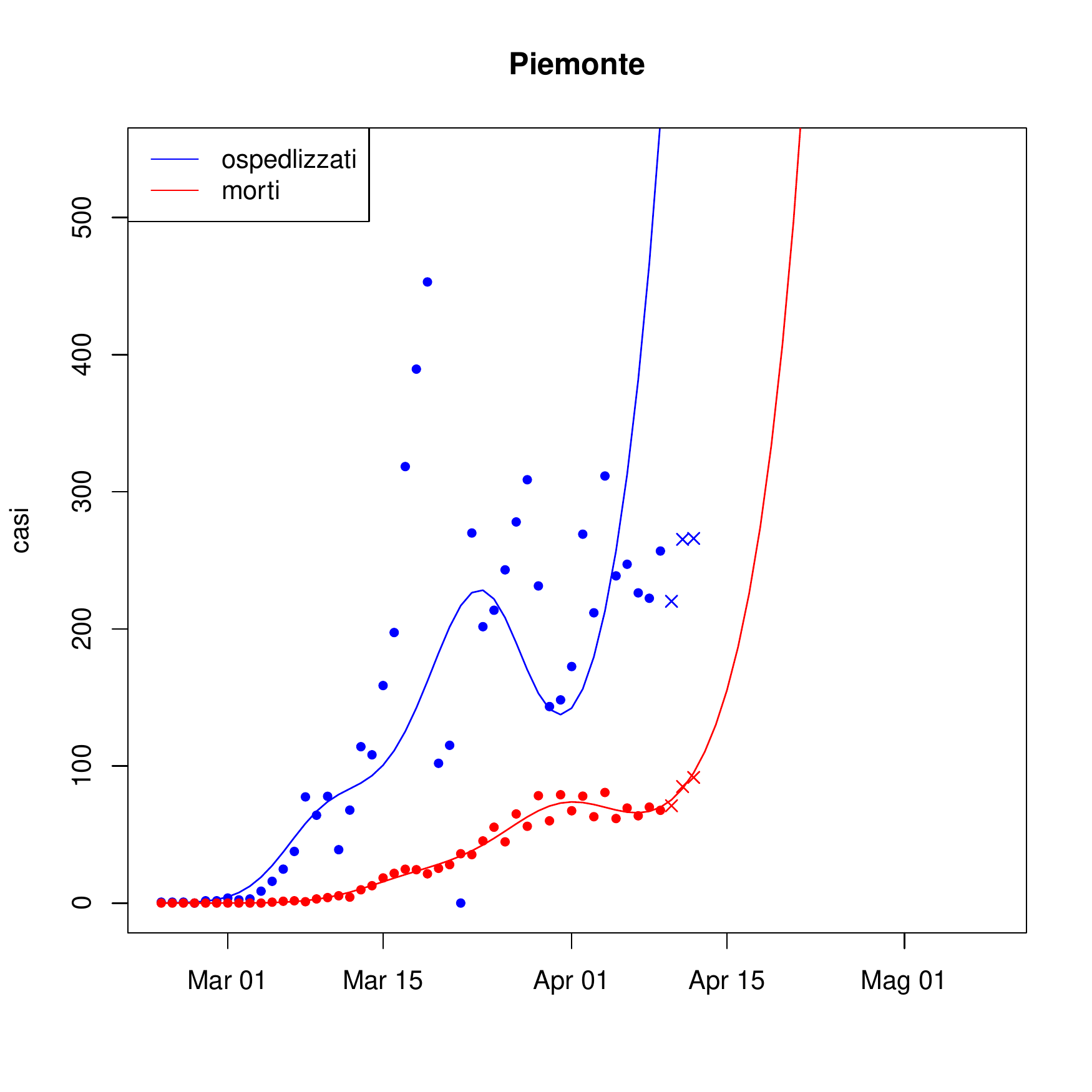}\ \includegraphics[width=6cm]{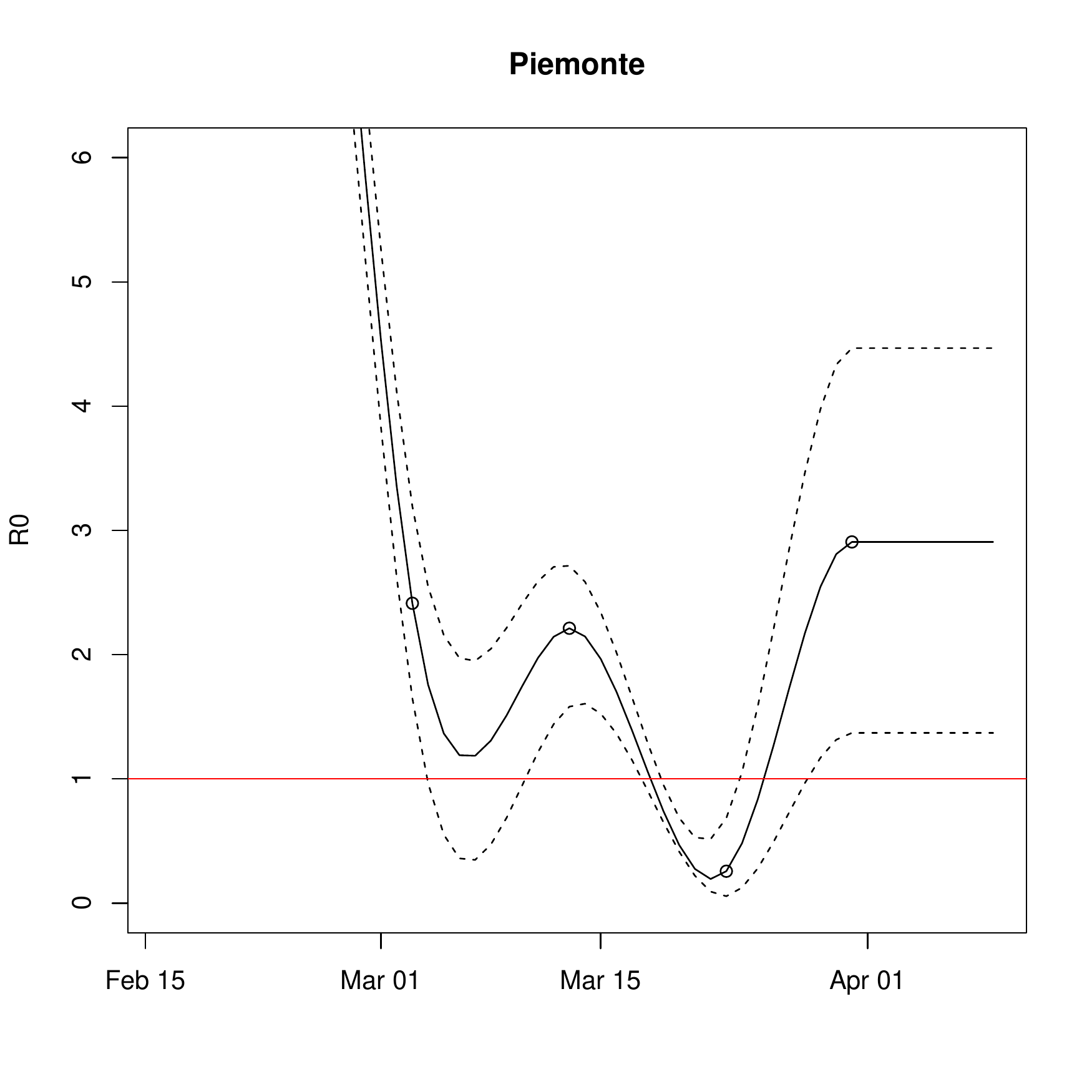}
\caption{Left: the points represent hospitalizations (blue) and deaths (red) from Piemonte. The solid curves are the expected values according to the best-fitting model to death data. Right: predicted values of $R_0$ with confidence intervals}
\label{fig:Piemonte}
\end{center}
\end{figure}

We fitted the model only to death data, although we stated that predictions that take into account also hospitalizations are more stable, because hospitalization data seemed unreliable especially for Piemonte (see Fig.~\ref{fig:Piemonte}).

\begin{figure}[htb]
\begin{center}
\includegraphics[width=6cm]{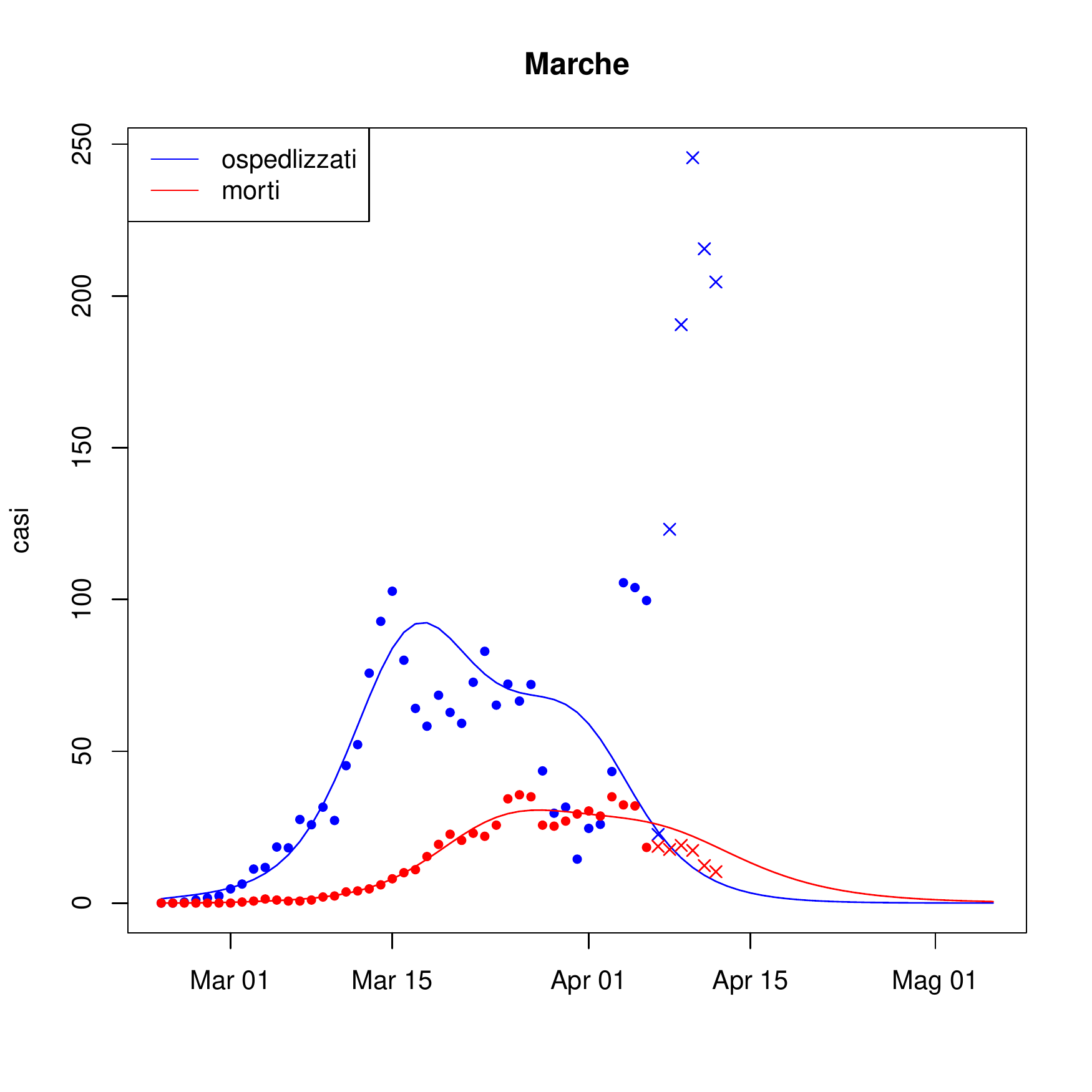}\ \includegraphics[width=6cm]{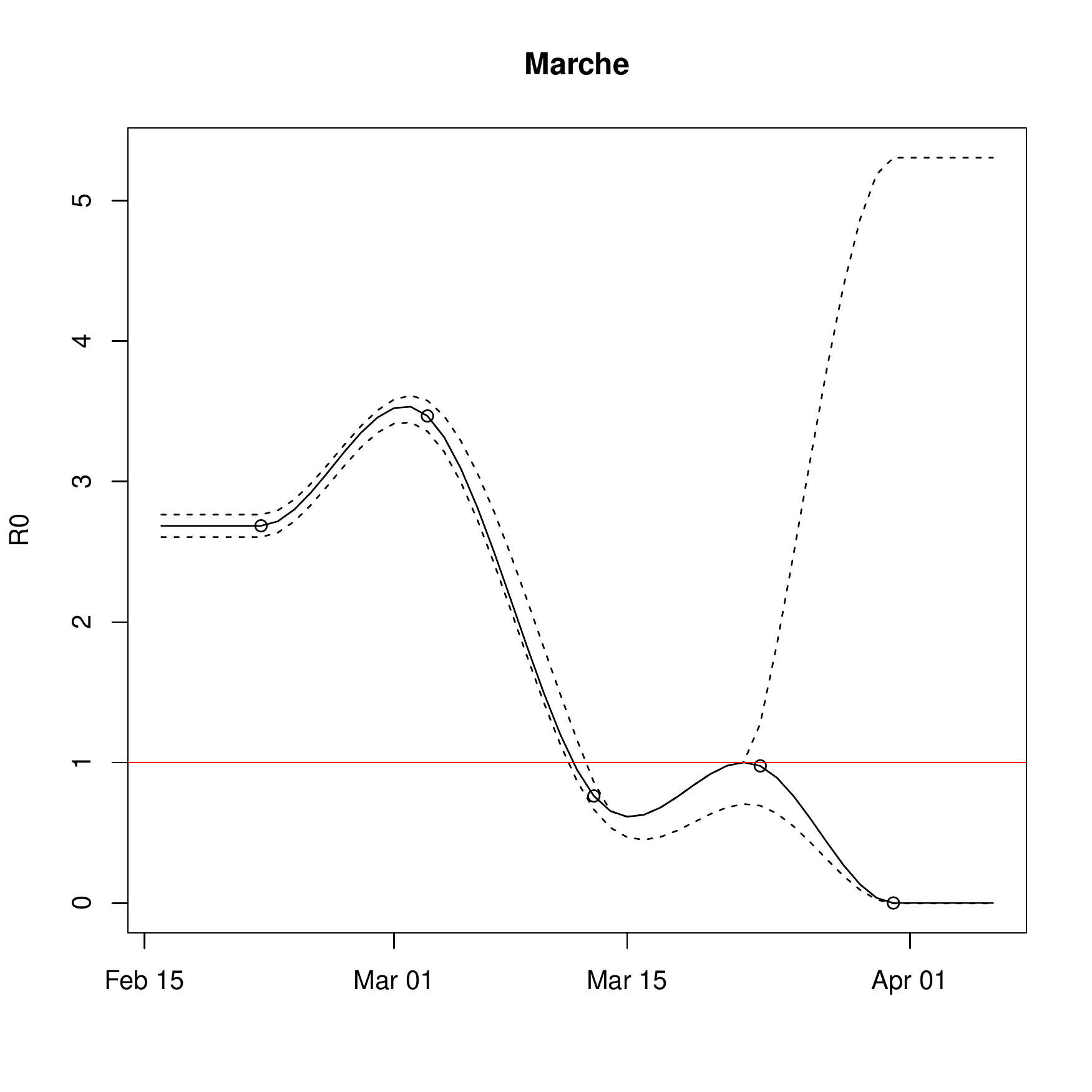}
\caption{Left: the points represent hospitalizations (blue) and deaths (red) from Marche. The solid curves are the expected values according to the best-fitting model to death data. Right: predicted values of $R_0$ with confidence intervals}
\label{fig:Marche}
\end{center}
\end{figure}

\section{Discussion}
The main results obtained from this analysis are an estimate of the curve of new infections over time, as well as an estimate of $R_0$. The estimates are valid also up to a few days before the final date of available data, although clearly with wide confidence interval. The estimates obtained for $R_0$ appear to be consistently below 1 since the second part of March in the whole Italy and in most regions, although there may be exceptions (some discussed below). 

A recent publication \cite{Riccardo2020} from the ISS-group has provided daily estimates for the effective reproduction number $R_t$ for 6 regions up to March 12, using much more detailed data. For the periods and regions when we could compare the two estimates (that actually are based on different assumptions), the trends are quite similar, although our estimates obtained for the early epidemic phase are higher, arriving up to 4. 

The method presented here, which is essentially a back-calculation, is based on simple aggregated data, and can be used on the most recent data, providing  quickly estimates of the current trends, estimates that can then be improved as more detailed data are available and analysed. Thus, we believe that the method, although sub-optimal relative to an analysis based on more detailed data (including for instance dates of symptom onset, and age structure of cases), can be helpful for public health to provide quick and rough estimates of the trends in new infections.

As mentioned in the beginning, the method is conceptually similar to what has been used by Flaxman \textit{et al.} \cite{Flaxman2020} for several European countries including Italy. Beyond technical differences (Flaxman \textit{et al.} \cite{Flaxman2020} used a Bayesian hierarchical model, calibrated to several European countries, which is presumably more robust than our approach), the two main differences from \cite{Flaxman2020} are the following ones: Flaxman \textit{et al.} \cite{Flaxman2020} fitted only data on deaths from which, through an indirect method, they were able to infer also the current number of infectives; secondly, in \cite{Flaxman2020} it is assumed that the contact rate is a piecewise-constant function that changed every time a new intervention was established by the government, while we let the contact rate be a smooth function. While certainly each government intervention imposed a discontinuity in behaviour, we believe that using spline functions gives more flexibility, and allows to estimate changes in adherence, due also to very frequent national and regional decrees that adjusted the exact level of restrictions.

The estimates of $R_0$ show a marked decrease already by the end of February, although a strong `lock-down' was in place only in a very small part of the country (around 50,000 inhabitants) and some closures (schools, movie-theatres\ldots) were in place only in a few regions (those considered here); it is possible that spontaneous changes in behaviour \cite{Manfredi_bk} helped in decreasing $R_0$, although still quite larger than 1.\\
 It can be noticed that our estimates of $R_0$ go consistently below 1 around March 20, more or less in correspondence of the government decree that ordered the closure of all non-essential working places.
  However, one has to consider that infections, when looked  at this aggregated level, will always show a certain inertia: as `lock-down' goes in place, new infections in the community will strongly decrease, but those that had already been infected will continue transmitting in households. It is then possible that already the previous measures could  bring $R_0$ below 1 if left in place long enough; more detailed data and models would be needed to solve the issue. Transmission within households could also explain the apparent rebounding of $R_0$ estimated for Emilia-Romagna near the beginning of March (Fig.~\ref{fig:ref_R0}c).

Using data on both hospitalizations and deaths  appears, as shown in Figures \ref{fig:weigh_fit} and \ref{fig:weigh_R0},  to yield much stabler estimates than using data from deaths alone; thus, we tend not to give too much weigh to the most recent estimate of $R_0 > 1$ obtained for the Veneto region, when only data on deaths are used; indeed in that case the profile likelihood in $\beta_5$ was not monotonic and, although the package \texttt{bbmle} still computes confidence intervals, they do not appear very reliable.  

As already stated, we did not consider data on daily positive tests. In order to use the number of positive tests, we believe if would be needed to model how they depend on the number of performed tests, and this seems rather difficult. We concentrated instead on data on hospitalizations and deaths, as they should be less dependent on policy choices. It is certainly possible that, in the moments and place where the sanitary system was under stress, requirements for  hospitalization became more stringent; in this sense, hospitalizations
may not be purely depend on the health of the patients, but also on the sanitary context.\\
Furthermore, a recent analysis of mortality rates in March suggests that the number of deaths related to COVID-19 has been higher than the numbers reported, at least in some areas of Lombardy region. Thus data on both deaths and hospitalizations may contain biases, but they are best data publicly available. Data on daily admissions to ICUs could have been useful, but we were not able to extract the information from the dataset.\\
We do not attempt to give an estimate of the actual number of infected individuals, but only of the curve, up to a multiplicative constant. In principle, one could use the observed death rates \cite{Riccardo2020} together with estimates of the `true' death rates to extrapolate the number of infections compatible with the observed deaths as in Flaxman \textit{et al.} \cite{Flaxman2020}

Although it is mainly a technical point, our analysis allows to discriminate among the different proposals for the distributions of times to death and hospitalizations. As can be seen from Table \ref{tab:uno}, the model chooses the hospitalization kernel that gives more weigh to longer delays (red curve in Fig.~\ref{fig:f1}a) and the  death kernel that gives more weigh to shorter delays (black curve in Fig.~\ref{fig:f1}b).\\
The selected curves are such that the mean time from symptom onset to hospitalization is 5.12 days (median 4.1) \cite{Pellis2020}, while the mean time from symptom onset to death is 15.2 days (median 13.8) \cite{Linton2020}. The data published by the Italian Health Institute \cite{ISS_report_8}
 estimate a median tim from symptom onset to hospitalization of 5 days, and from symptom onset to death of 10 days. This discrepancy between the estimates we used and Italian data might explain why the distance between the peak time of hospitalizations and of deaths is generally shorter in data than in our model (Fig.~\ref{fig:ref_fit}); the exception is the Veneto region (Fig.~\ref{fig:ref_fit}d), where both peaks are well estimated by the model, possibly because there the distribution of times are closer to estimates in the literature. One has to take into account that our analysis is based on reporting dates; hence, the delays between hospitalization or death and their reporting have also to be considered.\\ 
 Clearly, results depend in an essential way on good estimates of delay functions, and on the assumption they they are constant in time and among the regions. The method can be easily adapted to delays changing in time, as long as data are available to estimate this. 

We performed an analysis on how the results depended on some implementation details (such as the number of knots in the splines describing the contact rates) or the starting day of the simulation or the final day of data that are fitted. As can be seen (Fig.~\ref{fig:knots}), the number of knots changes the results very little; similarly changing the final date of fitting (Fig.~\ref{fig:final}) between April 3, 6 or 9 changes the estimates only for the very final part of fitting; moreover, the differences in the estimated final $R_0$ are minor, although extrapolating the simulations 30 days ahead they would bring quite a difference in the predicted number of hospitalizations.\\
Looking at the crosses (data not used in the fits) in all Figures, one can get an idea of the short-term predictive capabilities of the model.\\
One may notice that for a few days after April 6 the estimated number of hospitalizations have a large increase, that appears inconsistent with the model and the rest of the curve. We found out that the number of reported discharged patients in the four days between April 8 and 11 was over 2,000 per day, while previously it averaged around 1,000 a day with a peak around 1,500, and the number decreased again from April 12 onwards. Since the reported number of hospitalized patients did not decrease sharply, formula \eqref{data} estimates a peak of new hospitalizations between April 8 and 11 that is very far from model predictions. \\
We can think of three explanations: perhaps data reporting was not precise, and several patients reported as discharged between April 8 and 11 had actually been discharged previously, so that the correct hospitalization curve would have a less sharp decrease after the peak around March 20-25, and no peak between April 8 and 11. 
Another possibility is that a large number of patients was actually discharged between April 8 and 11, freeing hospital space and thus allowing to hospitalize patients previously in home isolation; in this case, the hospitalization peak of April 8-11 would have occurred, but because of procedures of the health system. Finally, the hospitalization peak might reflect a peak in new infections occured a few days earlier for unknown reasons. Probably the health authorities are aware of the hospital situation, but we think  that simple graphs like the ones presented here help in identifying patterns needing an explanation.

Finally, one can look at differences among the regions. Results from fitting to data from the whole Italy or from Lombardy are very similar, as expected, since around 40\% of the cases in Italy are from Lombardy. Veneto region presents curves with a much sharper peak, and generally the model fits very well data from Veneto; this is probably due to the rather homogeneous spread of infection within the region, and to a good quality of data acquisition and reporting. Emilia-Romagna is somewhat interemediate in this respect, possibly because of the presence of two infection foci at the ends of the region (Piacenza and Rimini). 
Data at the province level (NUTS 3) could provide a more accurate picture (although the smaller numbers would amplify the effect of noise in data), but unfortunately deaths and hospitalizations are not reported at that level.

A possible exception to the overall trends is given by the Piemonte region. There the infection started somewhat later than in the other regions in the North, but by now it has reported the third highest number of cases and deaths.\\
Despite our regularization procedure, the hospitalization data vary so widely from one day to the next one (Fig.~\ref{fig:Piemonte}) that we did not succeed in fitting the model to those data, so that the fitting procedure was applied only to data on deaths, that, as already discussed, provide less stable estimates, especially for recent dates. 
With that caveat, recent estimates of $R_0$ for Piemonte have been consistently above 1, and this corresponds also to the visual impression from data.\\
This is a case where the model fit is not optimal, but may raise an alarm bell, suggesting that more detailed analysis are recommended for that region, in order to better understand the situation.

\begin{acknowledgements}
We thank Ilaria Dorigatti, Alberto D'Onofrio, Giorgio Guzzetta, Piero Manfredi, Lorenzo Pellis, Stefania Salmaso and Gianpaolo Scalia Tomba for useful comments on previous drafts of the manuscript.
\end{acknowledgements}
\small
\bibliography{bibliography_file.bib}

 \end{document}